\documentclass[article]{aa}   

\usepackage{natbib}
\usepackage{graphicx}
\usepackage{txfonts}
\usepackage{array}
\usepackage{subfigure} 
\usepackage[skip=5pt]{caption} 
\usepackage{aalongtable}

\begin{document}

\title{Water ice deuteration: a tracer of the chemical history of protostars}


\author{V. Taquet$^1$ \and P. S. Peters$^{1,2}$ \and C. Kahane$^1$ \and C. Ceccarelli$^1$ \and A. L\'opez-Sepulcre$^1$ \and C. Toubin$^2$ \and D. Duflot$^2$ \and L. Wiesenfeld$^1$}

\offprints{V. Taquet: vianney.taquet@obs.ujf-grenoble.fr}

\institute{$^1$ UJF-Grenoble 1 / CNRS-INSU, Institut de Plan\'{e}tologie et d\textquoteright Astrophysique de Grenoble (IPAG) UMR 5274, Grenoble, F-38041, France \\
$^2$ Laboratoire de Physique des Lasers, Atomes et Molecules (PhLAM), UMR CNRS 8523, Universite Lille 1, Villeneuve d'Ascq Cedex 59655, France}

\date{Received / Accepted}

\titlerunning{Water ice deuteration: a tracer of the chemical history of protostars}
\authorrunning{V. Taquet et al.}

\abstract
{Millimetric observations have measured high degrees of molecular deuteration in several species seen around low-mass protostars. The \textit{Herschel Space Telescope}, launched in 2009, is now providing new measures of the deuterium fractionation of water, the main constituent of interstellar ices.}
{We aim at theoretically studying the formation and the deuteration of water, which is believed to be formed on interstellar grain surfaces in molecular clouds.}
{We used our gas-grain astrochemical model GRAINOBLE, which considers the multilayer formation of interstellar ices. We varied several input parameters to study their impact on water deuteration. We included the treatment of ortho- and para-states of key species, including H$_2$, which affects the deuterium fractionation of all molecules. The model also includes relevant laboratory and theoretical works on the water formation and deuteration on grain surfaces. In particular, we computed the transmission probabilities of surface reactions using the Eckart model, and we considered ice photodissociation following molecular dynamics simulations.}
{The use of a multilayer approach allowed us to study the influence of various parameters on the abundance and the deuteration of water. Deuteration of water is found to be very sensitive to the ortho-to-para ratio of H$_2$ and to the total density, but it also depends on the gas/grain temperatures and the visual extinction of the cloud. Since the deuteration is very sensitive to the physical conditions, the comparison with sub-millimetric observation towards the low-mass protostar IRAS 16293 allows us to suggest that water ice is formed together with CO$_2$ in molecular clouds with limited density, whilst formaldehyde and methanol are mainly formed in a later phase, where the condensation becomes denser and colder.}
{}
\maketitle

\section{Introduction}

Understanding the formation of water is crucial, not only because of its primordial importance for life on Earth, but also because it is thought to be one of the most abundant oxygen-bearing species and also one of the main gas coolants \citep{Ceccarelli1996, Kaufman1996, vanDishoeck2011}. 
Interstellar water is believed to be formed mainly via three main mechanisms: 1) cold gas-phase chemistry, starting from the ionization of H$_2$ by cosmic rays, eventually leading to H$_3$O$^+$ via ion-neutral reactions that then recombine with electrons to form H$_2$O \citep{Bates1986, Hollenbach2009}; 2) on the surface of interstellar dust particles, via the hydrogenation of accreted atomic and molecular oxygen occurring at cold temperatures \citep{Tielens1982, Cuppen2007, Miyauchi2008}; 3) warm gas chemistry, initiated by a few endothermic reactions involving H$_2$ in warm gas ($T > 250$ K) \citep{Ceccarelli1996, Kaufman1996}. 

The advent of space telescopes, combined with ground-based observatories, has allowed astronomers to observe vapours and ices of water in several phases of star formation. 
Water vapour is present in cold molecular clouds and prestellar cores but only with low abundances \citep[$X_{gas}$(H$_2$O) $\sim 10^{-8}-10^{-9}$ relative to H nuclei, see][]{Bergin2002, Klotz2008, Caselli2010}. In cold clouds, water is likely condensed in ices \citep[$X_{ice}$(H$_2$O) $\sim 5 \times 10^{-5} - 10^{-4}$,][]{Whittet1991, Pontoppidan2004}. 
Hot corinos and outflows of low-mass Class 0 protostars show higher abundances of gas phase water, with abundances of a few $10^{-6}$ in hot corinos \citep{Ceccarelli2000, Coutens2012, Kristensen2012}, whilst protostar outflows show higher abundances up to a few $10^{-5}$ \citep{Liseau1996, Lefloch2010, Kristensen2010, Kristensen2012}.  Water ice has also been observed towards cold protostellar envelopes with similar abundances to molecular clouds \citep[$\sim 10^{-4}$,][]{Pontoppidan2004, Boogert2008}.
Recent infrared observations have also shown the presence of water in protoplanetary disks in different states: water ice \citep{Terada2007} and hot and cold water vapour \citep[][with abundances of $10^{-4}$ and lower than $10^{-7}$, respectively]{Carr2008, Hogerheijde2011}. Analysis of debris disks show that dust particles are covered by ice mixtures \citep[e.g.][]{Li1998, Lebreton2012}. 

To summarise, it is now accepted that water is present during all phases of the star formation process. However, its evolution from molecular cloud to planetary system still remains unclear. The deuterium fractionation can help us constrain its formation and its evolution. First, it can probe the formation pathways of water observed in the early stages because of its sensitivity to the physical conditions. Second, it allows us to investigate its reprocessing in protoplanetary disks, and eventually to determine whether water on Earth has an interstellar origin. Comparing the HDO/H$_2$O ratio in comets and Earth is, for example, important for evaluating the possible contribution of comets for transferring water in Earth's oceans \citep{Owen1995}. Recent Herschel observations have reported a D/H ratio of water (0.014 \%) in the Jupiter family comet 103P/Hartley2 originating in the Kuiper Belt, very similar to the value for the Earth's oceans, supporting the hypothesis that a part of water comes from comets \citep{Hartogh2011}. 

\begin{figure}[htp]
\centering
\includegraphics[width=88mm]{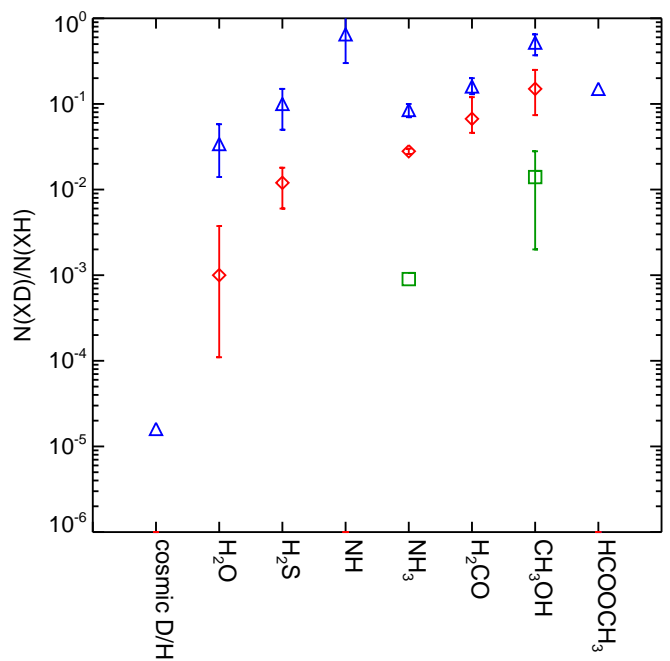}
\caption{Deuterium fractionation of several molecules assumed to be partly (or mainly) formed on interstellar grain surfaces and observed around low-mass Class 0 protostars. Blue triangles: simple deuteration; red diamonds: double deuteration; green squares: triple deuteration.}

\tablebib{ \\
\tablefoottext{a}{Cosmic D/H: \citet{Linsky2003}} \\
\tablefoottext{b}{H2O: HDO/H$_2$O by \citet{Coutens2012}, D$_2$O/H$_2$O by \citet{Vastel2010} towards IRAS 16293.} \\
 \tablefoottext{c}{H2S: HDS/H$_2$S by \citet{vanDishoeck1995} towards IRAS 16293, D$_2$S/H$_2$S by \citet{Vastel2003} towards IRAS4A.}\\
\tablefoottext{d}{NH: ND/NH by \citet{Bacmann2010} towards IRAS 16293.}\\
\tablefoottext{e}{NH3: NH$_2$D/NH$_3$ by \citet{vanDishoeck1995} towards IRAS 16293; NHD$_2$/NH$_3$ by \citet{Loinard2001} towards IRAS 16293, and ND$_3$/NH$_3$ by \citet{vanderTak2002} towards IRAS 4A.}\\
\tablefoottext{f}{H2CO: HDCO/H$_2$CO by \citet{vanDishoeck1995} towards IRAS 16293 and by \citet{Parise2006} towards seven low-mass protostars; D$_2$CO/H$_2$CO by \citet{Ceccarelli1998} and \citet{Ceccarelli2001} towards IRAS 16293 and by \citet{Parise2006} and \citet{Roberts2007} towards a sample of low-mass protostars.}\\
\tablefoottext{g}{CH3OH: CH$_2$DOH/CH$_3$OH and CHD$_2$OH/CH$_3$OH by \citet{Parise2006} towards seven low-mass protostars, CD$_3$OH/CH$_3$OH by \citet{Parise2004} towards IRAS 16293.}\\
\tablefoottext{h}{HCOOCH3: DCOOCH$_3$/HCOOCH$_3$ by \citet{Demyk2010} towards IRAS 16293.}\\
}
\label{deut_molecules}
\end{figure}

The HDO/H$_2$O ratio has recently been evaluated in the gas phase of low-mass Class 0 protostar envelopes with values varying from less than $10^{-4}$ in NGC1333-IRAS4B \citep{Jorgensen2010} to more than $10^{-2}$ in NGC1333-IRAS2A \citep{Liu2011}. The low-mass Class 0 protostar IRAS 16293-2422 seems to have the most reliable value since the main isotopologue (via H$_2^{18}$O) and its simply and doubly deuterated isotopologues have been observed several times via ground-based and space telescopes \citep{Ceccarelli2000, Parise2005, Butner2007, Vastel2010, Coutens2012}. The most recent work by \citet{Coutens2012} reports an HDO/H$_2$O ratio of $\sim 3$ \% in the hot corino, $\sim 0.5$ \% in the cold envelope, and $\sim 5$ \% in the photodesorption layer of the foreground cloud. Observations of D$_2$O by \citet{Butner2007} and \citet{Vastel2010} give a D$_2$O/H$_2$O ratio of $\sim 10^{-3}$. 
Using the 4.1 $\mu$m OD stretch band, solid HDO has been observed towards a sample of low- and high-mass stars by \citep{Dartois2003} and \citet{Parise2003}. These observations provide an upper limit for the solid abundance ratio between 0.2 and 2 \%. 
Although these values are much higher than the cosmic elemental abundance of deuterium \citep[$ 1.5 \times 10^{-5}$][]{Linsky2003}, water seems to be less deuterated than the other molecules also mainly formed on grain surfaces, such as formaldehyde and methanol. Figure \ref{deut_molecules} graphically shows that water possesses a lower level of deuteration, with a mean HDO/H$_2$O ratio of $\sim 3$ \%, whereas other molecules have a fractionation of their singly deuterated isotopologue higher than 10 \%. 

H$_2$O, but also H$_2$S, H$_2$CO and CH$_3$OH (and HCOOCH$_3$) are thought to be mainly formed on grain surfaces via the hydrogenation of accreted O (or O$_2$), S, and CO from the gas phase. 
Cold gas-phase chemistry produces water vapour with low abundances \citep[a few $10^{-7}$;][]{Bergin2000}, whilst endothermic reactions producing warm water are efficient at $T > 250$ K. In contrast, water ice desorbs into the gas phase in the envelope of Class 0 protostars in large quantities at $T \sim 100$ K. Therefore, the deuteration of water measured by millimetric observations likely reflects the deuteration of its icy precursor governed by the accretion of gas phase H and D atoms. Indeed, the timescale needed to significantly alter the deuteration after evaporation in warm gas is longer than the typical age of Class 0 protostars \citep[$\sim 10^5$ versus $\sim 10^4$ yr,][]{Charnley1997, Andre2000}. 
\citet{Roberts2003} showed that the gas phase atomic D/H ratio increases with the CO depletion. Molecules that are formed in the earlier stages of star formation, when the CO depletion is low, would, therefore, show lower deuteration. Based on our astrochemical model GRAINOBLE \citep[][hereafter TCK12a]{Taquet2012a}, we theoretically confirmed this hypothesis by successfully reproducing the observed formaldehyde and methanol deuterations \citep[][hereafter TCK12b]{Taquet2012b}. The difference in deuteration between these two molecules is explained by the earlier formation of formaldehyde compared to methanol, when the D/H is lower. We also demonstrated the necessity to introduce the abstraction reactions experimentally shown by \citet{Nagaoka2007} and \citet{Hidaka2009}. Similarly, \citet{Cazaux2011} showed the possibility that water is formed through reactions involving H$_2$. According to these authors, instead of reflecting the atomic D/H ratio, water deuteration should scale with the gas phase HD/H$_2$ ratio ($\sim 10^{-5}$) at low temperatures.

The linear relationship between the observed column density of water ice and the visual extinction, above the threshold of $A_V \sim 3.2$ mag found by \citet{Whittet1988}, suggests that water ice starts to form significantly at low visual extinctions. 
Other solid compounds are believed to form along with water ice. The linear relationship between carbon dioxide and water column densities, with a column density ratio $N$(CO$_2$)/$N$(H$_2$O) of 18 \%, suggests that these two molecules form in parallel  \citep{Whittet2007}. This conclusion is supported by comparisons between observed and laboratory band profiles showing that CO$_2$ is mainly located in a polar water-rich mixture whilst a non-polar component (pure CO$_2$, or CO:CO$_2$) exist in very low quantities \citep{Gerakines1999, Pontoppidan2008}.
Other solid organic compounds, such as CO or CH$_3$OH, have also been observed in large quantities ($\sim 30$ \% for CO, and up to $20 \%$ for methanol) but at higher visual extinctions, above $A_V$ thresholds of 8-9 and 15 mag respectively \citep{Whittet2007, Whittet2011}. 
Consequently, one needs to include the formation of all other solid molecules in order to correctly study the formation of deuterated water ice.

In this article, we extend our study to the formation and the deuteration of the interstellar water ice formed in molecular clouds and in prestellar cores, using the multilayer GRAINOBLE model. 
Our goal is to explore the influence of the physical conditions (representative of different typical cloud stages) on ice formation and water deuteration. This is the first time that such a systematic study has been done. Besides that, it includes the crucial influence of the H$_2$ ortho/para ratio on the water deuteration.
The astrochemical model is presented in Sect. \ref{model}. In Sect. \ref{results}, we study the formation of typical grain mantles and show the influence of several physical and chemical parameters on the deuteration of water ice. 
In each section, we summarise the main ideas with headings and concluding remarks.
In Sect. \ref{comp_models}, we compare our with previous model predictions and in Sect. \ref{sec: discussion} with published observations of water, formaldehyde, and methanol.

\section{Modelling} \label{model}

\subsection{Overview of the GRAINOBLE model}

GRAINOBLE is a gas-grain astrochemical model based on the rate equations approach introduced by \citet{Hasegawa1992}. A detailed presentation of GRAINOBLE can be found in TCK12a. 
Briefly, GRAINOBLE couples gas-phase and grain-surface chemistry. In total, our chemical network consists of 341 (gaseous and solid) species and 3860 (gas phase and grain surface) reactions. The gas phase chemistry is described in detail in section \ref{gas_phase}. 
The grain surface chemistry processes are the following: \\
i) The accretion of gas phase species onto the grain surfaces, assumed to be spherical. \\
ii) The diffusion of adsorbed species via thermal hopping. \\
iii) The reaction between two particles via the Langmuir-Hinshelwood mechanism, once they meet in the same site. The reaction rate is the product of the number of times that the two reactants meet each other and the transmission probability $P_r$ of reaction. \\
iv) The desorption of adsorbed species into the gas phase via several processes: \\
- thermal desorption; \\ 
- cosmic-ray induced heating of grains; \\ 
- chemical desorption caused by the energy release of exothermic reactions. \\ 
This last process is an upgrade with respect to TCK12a. Following \citet{Garrod2007}, we assume a value of 0.012 for the factor $a$ (the ratio of the surface-molecule bond frequency to the frequency at which energy is lost to the grain surface) since it seems to be the most consistent value given by molecular dynamics simulations \citep{Kroes2005}. \\
In addition, in the present work, we take the effect of the UV photolysis on the ices into account (see section \ref{section: photolysis}), because this is important for the formation of H$_2$O at low visual extinctions.
As suggested by laboratory experiments, cold mantle bulks are mostly inert \citep[see][]{Watanabe2004, Fuchs2009, Ioppolo2010}. Therefore, we follow the formation of grain mantles with a multilayer approach in which the outermost only layer is reactive, whilst the mantle bulk remains inert (see TCK12a for more details).

\subsection{Gas-phase chemical network} \label{gas_phase}

We consider the gas-phase chemical network from the KIDA database \citep{Wakelam2012} for seven elements: H, He, C, N, O, S, and Fe, giving a total of 258 gaseous species. Sulphur and iron are introduced to consistently study ion chemistry. Reactions involving atomic Fe and S play a significant role in the destruction of H$^+$ and H$_3^+$ whilst S$^+$, and to a lesser extent Fe$^+$, is believed to be one of the most abundant ions before the CO depletion \citep[see][]{Flower2005}.  
Relative to the KIDA network, we modify the rate coefficient of the cosmic-ray dissociation of H$_2$ yielding H+H. Following \citet{Dalgarno1999}, we assume $\gamma = 0.5$, where the rate $k_{diss}$ (in s$^{-1}$) of cosmic ray dissociation reactions is given by $k_{diss} = \gamma \zeta$ ($\zeta$ is the cosmic-ray ionization rate). 
Before ice formation, bare grains are considered. The recombination efficiency of atomic hydrogen and deuterium are assumed to be unity, following the theoretical works by \citet{Cazaux2004} and \citet{Cuppen2010b}, who considered chemisorption interactions.
The {initial elemental abundances in the gas phase} considered in this work are listed in Table \ref{elem_abu} and follows the work by \citet{Wakelam2010} and \citet{Linsky2003}. 
To model the water formation at low visual extinctions (see Introduction), we consider the depth-dependent self-shielding of H$_2$, HD, and CO, using the Meudon PDR code \citep{LePetit2002, Lepetit2006}.

{Since standard gas-phase models overpredict the O$_2$ abundance with respect to what is observed \citep[see][]{Goldsmith2011, Liseau2012} we also ran a grid of models by artificially decreasing the formation rate of O$_2$ by a factor of 10. We checked that the [O$_2$]/[O] abundance ratio remains indeed ten times lower than the ''standard" case throughout the whole calculation.}

Deuteration of water and other molecules formed on interstellar grains strongly depend on the gas phase abundances of H, D, H$_2$, HD, and D$_2$.
In turn, the abundance of these gaseous species, and more particularly D, depends mainly on the deuteration of H$_3^+$ which is function of the degree of CO and N$_2$ freeze-outs onto the grain surfaces \citep[see][]{Roberts2000, Roberts2003}. The deuterium gas phase chemical network is based on that of TCK12b. The major differences with respect to TCK12b are the inclusion of the ortho-to-para (hereinafter opr) H$_2$ ratio and reactions involving N$_2$.

Reactions between H$_3^+$ isotopologues and H$_2$, reducing the deuterium fractionation, are endothermic. Therefore, these reactions cannot occur in cold-cloud conditions if para-H$_2$ only is considered. 
However, H$_2$ is also believed to exist in the ortho spin state, higher in energy (170 K), since H$_2$ probably forms on grains with an ortho-to-para ratio (opr) of 3. Reactions between ortho-H$_2$ and ortho-H$_2$D$^+$ (ortho-HD$_2^+$, ortho-D$_3^+$) can reduce the degree of deuteration of H$_3^+$ significantly at low temperatures. \citet{Flower2006} have shown that the fractionation of H$_3^+$, hence, the abundance of atomic D are strongly reduced when the H$_2$ opr is higher than $10^{-4}$. 
Consequently, we enlarge the chemical network, relative to TCK12b, by considering the H$_3^+$-H$_2$ system whose new reaction rate coefficients have been computed by \citet{Hugo2009}. Ion-neutral reactions between H$_3^+$ isotopologues (including their different spin states) and CO or N$_2$, electronic recombinations, and recombinations on electronegative charged grains have also been included following \citet{Roberts2000, Roberts2003, Roberts2004, Walmsley2004} and \citet{Pagani2009}. 

The actual opr of H$_2$ in molecular clouds is still highly uncertain. The initial value of H$_2$ opr formed on grain surfaces is most likely 3, as recently confirmed by the experiment by \citet{Watanabe2010} conducted on amorphous solid water. Proton-exchange reactions in the gas phase would then convert ortho-H$_2$ to para-H$_2$, decreasing the opr of H$_2$ towards the Boltzmann value \citep[$\sim 3 \times 10^{-7}$ at 10 K, see][]{Flower2006}. 
Recent experimental studies have also demonstrated that H$_2$ undergoes a nuclear spin conversion from the ortho to the para spin state on amorphous solid water (ASW) \citep{Sugimoto2011, Chehrouri2011, Hama2012}.
The influence of the H$_2$ opr on absorption lines of formaldehyde has been observed by \citet{Troscompt2009} who deduce that the H$_2$ opr is much lower than 1. Indirect estimates of H$_2$ based on the comparison with chemical models suggest values of about $10^{-3}-10^{-2}$ \citep{Pagani2009, Dislaire2012}. Given the relative uncertainty in this value and its importance in the molecular deuteration process, in this work we assume the H$_2$ opr as a free parameter constant with time.

\begin{table}[htp]
\centering
\caption{{Initial elemental abundances in the gas phase} with respect to hydrogen nuclei {\citep[from][]{Linsky2003, Wakelam2010}}.}
\begin{tabular}{l c }
\hline
\hline
Species & Abundance \\
\hline
H$_2$ & 0.5 \\
HD & $1.6 \times 10^{-5}$ \\
He & 0.09 \\
C & $1.2 \times 10^{-4}$ \\
N & $7.6 \times 10^{-5}$ \\
O & $2.6 \times 10^{-4}$ \\
S & $8.0 \times 10^{-8}$ \\
Fe & $1.5 \times 10^{-8}$ \\
\hline
\end{tabular}
\label{elem_abu}
\end{table}

\subsection{Chemical network on grain surfaces}

\subsubsection{Formation and deuteration of water ice}

We consider a chemical network based on the work by \citet{Tielens1982} modified following the results of several recent experimental works as described below.
The simplest formation pathway towards solid water is the sequential hydrogenation of atomic oxygen:
\begin{center}
\begin{equation} \textrm{O} + \textrm{H} \rightarrow \textrm{OH} \label{O_H} \end{equation}
\begin{equation} \textrm{OH} + \textrm{H} \rightarrow \textrm{H}_2\textrm{O}. \label{OH_H} \end{equation}
\end{center}
This reaction channel was experimentally measured to occur in cold conditions, probably via barrierless reactions \citep{Hiraoka1998, Dulieu2010, Jing2011}. 

In addition, water can also be formed from different channels involving O$_2$ or O$_3$ as follows. 
First, water can be formed from molecular oxygen, either from the gas phase or formed on grains, following the reaction channels:
\begin{center}
\begin{equation} \textrm{O}_2 + \textrm{H} \rightarrow \textrm{HO}_2 \end{equation}
\begin{equation} \textrm{HO}_2 + \textrm{H} \rightarrow \textrm{H}_2\textrm{O}_2  \end{equation}
\begin{equation} \textrm{H}_2\textrm{O}_2 + \textrm{H} \rightarrow \textrm{H}_2\textrm{O} + \textrm{OH}. \label{H2O2_H} \end{equation}
\end{center}
These pathways have been experimentally demonstrated by \citet{Miyauchi2008, Ioppolo2010} and \citet{Cuppen2010a} at temperatures of 10 K. \citet{Miyauchi2008} also showed an isotope effect in the formation of water from hydrogen peroxide, implying the possibility of tunnelling through an activation barrier for this reaction.

Second, water can be formed from ozone on interstellar grains \citep{Cuppen2007,Taquet2012a} 
following the reaction
\begin{center}
\begin{equation} \textrm{O}_3 + \textrm{H} \rightarrow \textrm{O}_2 + \textrm{OH} \end{equation}
\end{center}
Then, O$_2$ and OH can continue to react to form water via the reactions described above. \citet{Mokrane2009} and \citet{Romanzin2011} experimentally showed the efficiency of this reaction by observing the presence of water (HDO, D$_2$O) after the irradiation of solid O$_3$ on water ice by H (D) atoms. We treat this reaction as barrierless.

With their microscopic Monte Carlo model, \citet{Cuppen2007} have concluded that molecular hydrogen plays a key role in the formation of water ice in molecular clouds. When hydrogen is mostly in its molecular form, water is formed at $\sim 70$\% by the reaction 
\begin{equation} \textrm{OH} + \textrm{H}_2 \rightarrow  \textrm{H}_2\textrm{O} + \textrm{H}. \label{OH_H2} \end{equation}
\citet{Oba2012} experimentally showed that HDO formation from OH is ten times less efficient than the formation of H$_2$O from OH, implying an isotope effect and therefore the possibility of tunnelling through an activation barrier in this reaction.

OH radicals can also recombine if the ice temperature is high enough (40 K in their experiment) to allow their mobility. \citet{Oba2011} experimentally determined branching ratios for the reactions
\begin{equation} \textrm{OH} + \textrm{OH} \rightarrow \textrm{H}_2\textrm{O} + \textrm{O} \end{equation}
\begin{equation} \textrm{OH} + \textrm{OH} \rightarrow \textrm{H}_2\textrm{O}_2 \end{equation}
of 0.2 and 0.8, respectively. 

Unlike \citet{Cazaux2010, Cazaux2011}, we do not include the O+H$_2$ reaction since it is unlikely to proceed at low temperatures (10-20 K) given its high endothermicity \citep[960 K;][]{Baulch1992}. \citet{Oba2012} confirmed that the co-deposition of cold O atoms with H$_2$ at 10 K does not result in the formation of water but only of O$_2$. 

We enlarge the water-formation network by including the deuterated counterparts of all reactions mentioned above. Due to the higher mass of deuterated species with respect to their main isotopologue, the reaction rates of barrierless reactions involving deuterated species are decreased. A careful treatment of the transmission probabilities (i.e. probability of tunnelling through the activation barrier) of all reactions possessing a barrier is described in section \ref{section: eckart}.

\subsubsection{Formation and deuteration of other ices}

The accretion of CO and O particles onto interstellar grains can lead to the formation of carbon dioxide (CO$_2$) mainly via three reaction channels \citep{Ruffle2001b}
\begin{equation} \textrm{HCO} + \textrm{O} \rightarrow \textrm{CO}_2 + \textrm{H} \label{CO2_1} \end{equation}
\begin{equation} \textrm{CO} + \textrm{O} \rightarrow  \textrm{CO}_2  \label{CO2_3} \end{equation}
\begin{equation} \textrm{CO} + \textrm{OH} \rightarrow \textrm{CO}_2 + \textrm{H}  \label{CO2_2} \end{equation}
Gas phase experiments have shown that reaction (\ref{CO2_1}) is barrierless \citep[e.g. ][]{Baulch2005} although it has never been studied on interstellar ice analogues so far. 

Reaction (\ref{CO2_3}) is thought to have a high activation energy \citep{Talbi2006}. Laboratory experiments have shown that solid carbon dioxide can be formed from this reaction with low efficiency, at least two orders of magnitude lower than astronomical observations \citep{Roser2001, Raut2011}.

Formation of solid carbon dioxide, from OH radicals and CO molecules, has been observed even at very low temperatures \citep[10-20 K,][]{Oba2010, Ioppolo2011, Noble2011}. However, the exact pathway of the formation of CO$_2$ is still uncertain. Following \citet{Oba2010} who observed a weak band attributed to the HOCO radical and other theoretical works, we propose a reaction pathway for the CO$_2$ formation from CO and OH, as described in Appendix A.2. 

The chemical network presented in TCK12b, with the same relative rates, is used to study the formation and the deuteration of formaldehyde and methanol.
We also consider the formation of deuterated methane and ammonia from barrierless addition reactions of solid atomic carbon and nitrogen. 
A full list of surface reactions considered in this work is presented in Appendix B.

\subsection{Eckart model and reaction probabilities} \label{section: eckart}

In previous gas-grain astrochemical models, the transmission probability of exothermic surface reactions has been approximated by the exponential portion of the quantum mechanical probability for tunnelling through a square potential barrier of width $a$. However, this approach has two main limitations: \\
i) it does not fit the potential energy profile of reactions correctly; \\
ii) the width $a$ is unknown, although in most astrochemical models it is arbitrarily fixed to $1 \AA$ \citep{Tielens1982, Hasegawa1992}. \citet{Garrod2011} reproduced the value of the transmission probability of a few reactions with square potential barriers and deduced a width of about 2$\AA$. However, the deduction of this width is based on poorly constrained activation barriers.

Since square potential barriers do not allow us to accurately estimate the values of the transmission probabilities, we compute the transmission probabilities of all the reactions using the Eckart model \citep{Eckart1930, Johnston1962}, which fits an approximate potential energy surface. A full description of the Eckart model, our quantum chemistry calculations and the computations of the transmission probabilities are given in Appendix A.

\begin{table*}[htp]
\centering
\caption{List of reactions having a barrier, with the forward, backward activation barriers, the imaginary frequency of the transition state, the transmission probabilities computed with the Eckart model and the square barrier method (using a barrier width of 1 $\AA$).}
\begin{tabular}{l c c c c c c c c c c c c c c}
\hline
\hline
Reactions	&		&		&		&		&		&		&	Probability type	&	V$_f$ (K)	&	V$_b$ (K)	&	$\nu_S$	&	$P_{r,Eckart}$	&	$P_{r,square}$	&	Reference	\\
\hline																											
OH	&	+	&	H$_2$	&	$\rightarrow$	&	H$_2$O	&	+	&	H	&	Eckart	&	2935	&	10209	&	-1293	&	4.07(-07)	&	1.49(-13)	&	1	\\
OH	&	+	&	HD	&	$\rightarrow$	&	H$_2$O	&	+	&	D	&	Eckart	&	2855	&	9396	&	-1259	&	3.62(-07)	&	7.91(-16)	&	1	\\
OH	&	+	&	HD	&	$\rightarrow$	&	HDO	&	+	&	H	&	Eckart	&	3051	&	10508	&	-970.4	&	1.00(-09)	&	2.44(-16)	&	1	\\
OH	&	+	&	D$_2$	&	$\rightarrow$	&	HDO	&	+	&	D	&	Eckart	&	3026	&	9556	&	-955.3	&	8.07(-10)	&	8.02(-18)	&	1	\\
OD	&	+	&	H$_2$	&	$\rightarrow$	&	HDO	&	+	&	H	&	Eckart	&	2789	&	10246	&	-1293	&	8.74(-07)	&	2.88(-13)	&	1	\\
OD	&	+	&	HD	&	$\rightarrow$	&	D$_2$O	&	+	&	H	&	Eckart	&	2900	&	10871	&	-970.1	&	2.81(-09)	&	5.2(-16)	&	1	\\
OD	&	+	&	HD	&	$\rightarrow$	&	HDO	&	+	&	D	&	Eckart	&	2703	&	9736	&	-1258	&	7.99(-07)	&	1.76(-15)	&	1	\\
OD	&	+	&	D$_2$	&	$\rightarrow$	&	D$_2$O	&	+	&	D	&	Eckart	&	2870	&	10338	&	-955.1	&	2.26(-09)	&	7.03(-18)	&	1	\\
\hline																											
H$_2$O$_2$	&	+	&	H	&	$\rightarrow$	&	H$_2$O	&	+	&	OH	&	Eckart	&	2508	&	36358	&	-1054	&	1.37(-07)	&	1.18(-08)	&	2	\\
H$_2$O$_2$	&	+	&	D	&	$\rightarrow$	&	HDO	&	+	&	OH	&	Eckart	&	2355	&	37118	&	-843.7	&	5.54(-09)	&	8.83(-12)	&	2	\\
HDO$_2$	&	+	&	H	&	$\rightarrow$	&	HDO	&	+	&	OH	&	Eckart	&	2523	&	36239	&	-1053	&	1.23(-07)	&	1.17(-08)	&	2	\\
HDO$_2$	&	+	&	H	&	$\rightarrow$	&	H$_2$O	&	+	&	OD	&	Eckart	&	2524	&	36063	&	-1053	&	1.22(-07)	&	1.17(-08)	&	2	\\
HDO$_2$	&	+	&	D	&	$\rightarrow$	&	HDO	&	+	&	OD	&	Eckart	&	2369	&	36822	&	-846.6	&	5.28(-09)	&	8.66(-12)	&	2	\\
HDO$_2$	&	+	&	D	&	$\rightarrow$	&	D$_2$O	&	+	&	OH	&	Eckart	&	2367	&	37023	&	-846.1	&	5.29(-09)	&	8.66(-12)	&	2	\\
D$_2$O$_2$	&	+	&	H	&	$\rightarrow$	&	HDO	&	+	&	OD	&	Eckart	&	2540	&	35938	&	-1052	&	1.08(-07)	&	1.17(-08)	&	2	\\
D$_2$O$_2$	&	+	&	D	&	$\rightarrow$	&	D$_2$O	&	+	&	OD	&	Eckart	&	2384	&	36721	&	-842.9	&	4.28(-09)	&	8.49(-12)	&	2	\\
\hline																											
CO	&	+	&	H	&	$\rightarrow$	&	HCO	&		&		&	Eckart	&	1979	&	8910	&	-793.6	&	1.92(-07)	&	1.83(-08)	&	3	\\
CO	&	+	&	D	&	$\rightarrow$	&	DCO	&		&		&	Experiments	&		&		&		&	1.92(-08)	&	1.83(-09)	&	4	\\
\hline																											
H$_2$CO	&	+	&	H	&	$\rightarrow$	&	CH$_3$O	&		&		&	Experiments	&		&		&		&	9.60(-08)	&	9.15(-09)	&	4	\\
H$_2$CO	&	+	&	D	&	$\rightarrow$	&	CH$_2$DO	&		&		&	Experiments	&		&		&		&	9.60(-09)	&	9.15(-10)	&	5	\\
H$_2$CO	&	+	&	D	&	$\rightarrow$	&	HCO	&	+	&	HD	&	Experiments	&		&		&		&	9.31(-08)	&	8.88(-09)	&	4	\\
H$_2$CO	&	+	&	D	&	$\rightarrow$	&	HDCO	&	+	&	H	&	Experiments	&		&		&		&	9.31(-08)	&	8.88(-09)	&	4	\\
HDCO	&	+	&	H	&	$\rightarrow$	&	CH$_2$DO	&		&		&	Experiments	&		&		&		&	1.11(-07)	&	1.06(-09)	&	4	\\
HDCO	&	+	&	D	&	$\rightarrow$	&	CHD$_2$O	&		&		&	Experiments	&		&		&		&	1.11(-07)	&	1.06(-09)	&	5	\\
HDCO	&	+	&	H	&	$\rightarrow$	&	HCO	&	+	&	HD	&	Experiments	&		&		&		&	1.54(-07)	&	1.46(-08)	&	5	\\
HDCO	&	+	&	D	&	$\rightarrow$	&	DCO	&	+	&	HD	&	Experiments	&		&		&		&	9.31(-08)	&	8.87(-09)	&	5	\\
HDCO	&	+	&	D	&	$\rightarrow$	&	D$_2$CO	&	+	&	H	&	Experiments	&		&		&		&	9.31(-08)	&	8.87(-09)	&	5	\\
D$_2$CO	&	+	&	H	&	$\rightarrow$	&	CH$_2$DO	&		&		&	Experiments	&		&		&		&	1.27(-07)	&	1.21(-08)	&	4	\\
D$_2$CO	&	+	&	D	&	$\rightarrow$	&	CD$_3$O	&		&		&	Experiments	&		&		&		&	1.27(-08)	&	1.21(-09)	&	5	\\
D$_2$CO	&	+	&	H	&	$\rightarrow$	&	DCO	&	+	&	HD	&	Experiments	&		&		&		&	7.30(-08)	&	6.95(-09)	&	4	\\
\hline																											
CH$_3$OH	&	+	&	D	&	$\rightarrow$	&	CH$_2$OH	&	+	&	HD	&	Experiments	&		&		&		&	2.88(-07)	&	2.75(-08)	&	4	\\
CH$_2$DOH	&	+	&	D	&	$\rightarrow$	&	CHDOH	&	+	&	HD	&	Experiments	&		&		&		&	1.92(-07)	&	1.83(-08)	&	4	\\
CHD$_2$OH	&	+	&	D	&	$\rightarrow$	&	CD$_2$OH	&	+	&	HD	&	Experiments	&		&		&		&	1.50(-07)	&	1.43(-08)	&	4	\\
CH$_3$OD	&	+	&	D	&	$\rightarrow$	&	CH$_2$OD	&	+	&	HD	&	Experiments	&		&		&		&	2.88(-07)	&	2.75(-08)	&	5	\\
CH$_2$DOD	&	+	&	D	&	$\rightarrow$	&	CHDOD	&	+	&	HD	&	Experiments	&		&		&		&	1.92(-07)	&	1.83(-08)	&	5	\\
CHD$_2$OD	&	+	&	D	&	$\rightarrow$	&	CD$_2$OD	&	+	&	HD	&	Experiments	&		&		&		&	1.50(-07)	&	1.43(-08)	&	5 \\
\hline
\end{tabular}
\tablefoot{The probability type refers to the method used for
  computing the probability. ``Experiments'' means that the
  transmission probability is deduced from the CO+H reaction by considering the relative rates experimentally measured by \citet{Hidaka2009} (see TCK12b for more details). }
\tablebib{1: \citet{Nguyen2011}, 2: This work. 3: \citet{Peters2012}, 4: \citet{Hidaka2009}, 5: \citet{Taquet2012b}.
}			
\label{reacs_eckart}
\end{table*}

Briefly, the formation of deuterated water includes two reaction channels which have an activation barrier: reactions (\ref{H2O2_H}) and (\ref{OH_H2}). 
Reaction (\ref{H2O2_H}) has been theoretically studied by \citet{Koussa2006} and \citet{Ellingson2007} but only for the main isotopologues. Therefore, quantum chemistry calculations are conducted for all the reactions (\ref{H2O2_H}) including deuterated isotopologues (see Appendix A for more details on the calculations). 
Concerning reaction (\ref{OH_H2}), the transmission probabilities of all the deuterated counterparts have been deduced from the theoretical work by \citet{Nguyen2011}. 

\begin{figure}[htp]
\centering
\includegraphics[width=88mm]{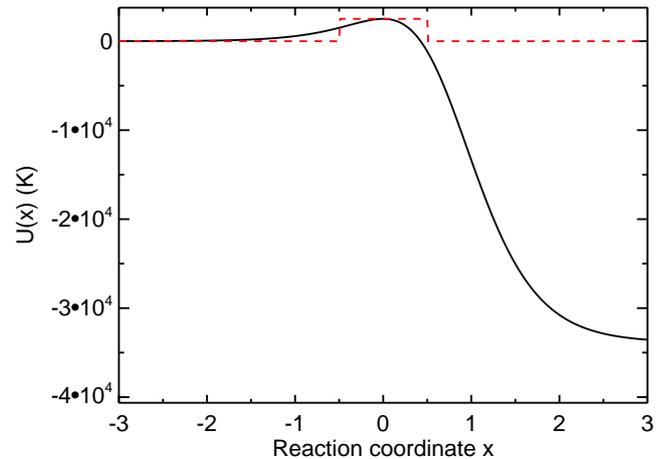}
\caption{Potential energy profile as function of the reaction coordinate of the H$_2$O$_2$ + H $\rightarrow$ H$_2$O + OH reaction computed from ab-initio calculations (black solid curve) and adopting a symmetric square barrier of a width of 1 $\AA$ (red dashed curve). See Appendix A for more details on calculations.}
\label{eckart_H2O2}
\end{figure}

Figure \ref{eckart_H2O2} shows the Eckart and the symmetric square potentials as a function of reaction coordinates for the reaction (\ref{H2O2_H}) computed in this work. As suggested previously, it can be seen that the Eckart potential is far to be symmetric. Furthermore, the reaction profile is thinner than the square barrier. Since the quantum tunnelling probability of transmission through this barrier depends on the area of the potential energy profile, the Eckart model gives higher probability of transmission than the square barrier ($1.4 \times 10^{-7}$  versus $1.2 \times 10^{-8}$).

Table \ref{reacs_eckart} lists the reactions with an activation barrier, as well as the input parameters needed for computing the transmission probabilities. It also compares the transmission probability computed with the Eckart model and with a symmetric square potential barrier of the same activation energy and adopting a width of 1 $\AA$ (the value commonly used in most gas-grain astrochemical models). The comparisons between the two approaches show that the assumption of a square barrier width of 1 $\AA$ tends to underestimate the reaction probabilities for all the reactions by up to seven orders of magnitude. 

The transmission probability of the CO+H reaction computed with the Eckart model is in good agreement with the range of values we deduced in TCK12a. In this latter work, we varied the transmission probability $P_r$ of this reaction and found that a transmission probability higher than $2 \times 10^{-7}$ was needed to reproduce the solid CH$_3$OH/CO ratio observed towards high-mass protostars.

\subsection{Photodissociation and photodesorption of ices} \label{section: photolysis}

In addition to Langmuir-Hinshelwood chemical reactions, we also consider the effect of FUV (6 - 13.6 eV) photons on ices, following the results of molecular dynamics (MD) simulations carried out by \citet{Andersson2006} and \citet{Andersson2008}. 
They showed that an amorphous ice absorbs UV photons in a 1-2 eV narrow band peaked at $\sim 8.5$ eV, in good agreement with experimental works by \citet{Kobayashi1983}, reaching a maximal absorption probability of $7 \times 10^{-3}$ per monolayer. We convolve the absorption spectrum with the emission spectrum of the interstellar radiation field (ISRF) deduced by \citet{Mathis1983} and with the emission spectrum of H$_2$ excited by the secondary electrons produced by the cosmic-rays ionization of H$_2$ (CRH2RF) computed by \citet{Gredel1987}. The absorption probabilities $P_{abs}$ of each monolayer integrated along the 6 - 13.6 eV band are therefore equal to $1.51 \times 10^{-3}$ for ISRF and $1.03 \times 10^{-3}$ for CRH2RF.
\citet{Andersson2008} also showed that the photofragments, OH and H, display different trajectories (desorption, trapping, or mobility on the surface) and they computed the yield of each trajectory as function of the monolayer. 
Subsequent experimental studies by \citet{Yabushita2006, Yabushita2009} and \citet{Hama2009, Hama2010} have confirmed the different photofragment trajectories revealed by MD simulations. However, our model only considers one chemically reactive layer (the outermost). We deduce the yield of each trajectory by averaging the yields computed by \citet{Andersson2008} on each layer. The list of trajectories as well as their averaged yield $Y$ is given in Table \ref{photoreactions}. The rate of each trajectory can be deduced, via the following equation
\begin{equation}
R_{photo,i} (s^{-1}) = Y_i \times F_{UV} \times \sigma(a_d) \times 5 \times P_{abs} \times 1 / N_s \label{eq_photo}
\end{equation}
where $F_{UV}$ (cm$^{-2}$ s$^{-1}$) is the photon flux of the 6 - 13.6 eV UV band which is absorbed by the amorphous ice, $\sigma(a_d)$ (cm$^2$) is the cross section of interstellar grains ($= \pi (a_d/2)^2$ where $a_d$ is the grain diameter), 5 refers to the absorption of the five outermost layers, and $N_s$ is the number of sites on the grain surface. 
{Owing to the lack of quantitative data on the photodesorption of atoms and on the photodissociation of hydrogenated (and deuterated) molecules other than water on ASW ice, we consider the same absorption probability for all atoms and molecules as for water. For atoms, the desorption probability upon absorption is assumed to be unity, whilst we consider the same outcome probabilities for hydrogenated molecules. }

\begin{table}[htp]
\centering
\caption{List of trajectories after H$_2$O photodissociation with their probability \citep[deduced from][see text]{Andersson2008}.}
\begin{tabular}{l c }
\hline
\hline
Outcome & Probability \\
\hline
H$_2$O$_{ice}$ $\rightarrow$ H$_{gas}$ + OH$_{ice}$ & 0.5 \\
H$_2$O$_{ice}$ $\rightarrow$ H$_{ice}$ + OH$_{ice}$ & 0.2 \\
H$_2$O$_{ice}$ $\rightarrow$ H$_{ice}$ + OH$_{gas}$ & $2.2 \times 10^{-3}$ \\
H$_2$O$_{ice}$ $\rightarrow$ H$_{gas}$ + OH$_{gas}$ & $6.8 \times 10^{-3}$ \\
H$_2$O$_{ice}$ $\rightarrow$ H$_2$O$_{gas}$ & $4.0 \times 10^{-3}$ \\
H$_2$O$_{ice}$ $\rightarrow$ H$_2$O$_{ice}$ & $0.28$ \\
\hline
\end{tabular}
\label{photoreactions}
\end{table}

We follow the experimental results by \citet{Fayolle2011} for wavelength-dependent CO photodesorption between 7.5 and 13 eV. 
The convolution of the CO photodesorption spectrum with the ISRF and CRH2RF fields gives integrated photodesorption yields of $1.2 \times 10^{-2}$ and $9.4 \times 10^{-3}$ photon$^{-1}$ molecule $^{-1}$, respectively. Owing to the lack of data on the wavelength-dependent photodesorption of other molecules, we consider the same photodesorption rates for O$_2$, and N$_2$.

\subsection{Binding energies} \label{binding}

Comparisons between the observed absorption 3 $\mu$m band of water and laboratory experiments have shown that grain mantles are mainly composed of high-density amorphous solid water (ASW) \citep{Smith1989, Jenniskens1995}. 
Therefore, binding energies of adsorbed species relative to ASW must be considered.

It is now accepted that light particles (H, D, H$_2$, HD, D$_2$) show a distribution of their binding energies relative to ASW, depending on the ice properties, the adsorption conditions, and the coverage of the accreted particles. 
\citet{Perets2005} have experimentally highlighted the influence of the ice density on the HD and D$_2$ binding energies. Molecules adsorbed on low-density amorphous ices (LDI) desorb following three desorption peaks that are at lower temperatures than the single observed desorption peak of molecules evaporating from a high-density amorphous ice (HDI). 
By depositing D$_2$ on ASW, \citet{Hornekaer2005} showed that D$_2$ is more efficiently bound to porous surfaces. Furthermore, binding energies follow broad distributions between 300 and 500 K and between 400 and 600 K for non-porous and porous ASW ices, respectively. 
\citet{Amiaud2006} studied the link between the binding energy distribution of D$_2$ with its coverage on a porous ASW ice and with ice temperature. More particularly, they showed that the D$_2$ binding energy decreases with H$_2$ coverage, from $\sim 700$ K to $\sim 350$ K whilst distributions broaden. 

The experimental results have been supported by theoretical MD calculations. For example, \citet{Hornekaer2005} and \citet{AlHalabi2007}  found that the binding energy distributions are essentially a consequence of the variation in the number of water molecules surrounding the adsorbed particle. Thus, the binding energy of H shows a broader distribution peaked at higher values on porous and irregular amorphous water ice than on a structured crystalline ice.

It is believed that deuterated species are more efficiently bound with ices than their main isotopologues because of their higher mass \citep{Tielens1983}. However, the difference in binding energy between H, H$_2$, and their deuterated counterparts still remains poorly constrained. Experiments by \citet{Perets2005} and \citet{Kristensen2011} have shown that distribution peaks of the H$_2$, HD, and D$_2$ binding energies are very close (less than 5 meV $\sim$ 60 K). These differences remain much smaller than the typical full-width-at-half-maximum of binding energy distributions shown by MD simulations \citep[$\sim$ 195 K for a amorphous ice][]{AlHalabi2007}, we, therefore, assume the same binding energy for H, D, H$_2$, HD, and D$_2$. We also verify a posteriori that a small difference of 50 K between these binding energies has a very limited influence on the deuteration of water.
 
To take the binding energy distribution of light adsorbed species on ices into account, we assume the binding energy of H, D, H$_2$, HD, and D$_2$ relative to ASW $E_{b,ASW}$ as a free parameter between 400 and 600 K. We consider constant binding energies of heavier species, following several experiments. Table \ref{table_energies} lists the binding energies of selected species. We assume that the deuterated species have the same binding energy as their main isotopologue.

\begin{table}[htp]
\centering
\caption{List of selected species and binding energies relative to amorphous solid water ice.}
\begin{tabular}{c c c}
\hline
\hline
Species & $E_b$ (K)  \\
\hline
H & 400 - 600 \tablefootmark{a} \\
H$_2$ & 400 - 600 \tablefootmark{a} \\
C & 800 \tablefootmark{b} \\
N & 800 \tablefootmark{b} \\
O & 800 \tablefootmark{b} \\
CO & 1150 \tablefootmark{c} \\
CO$_2$ & 2690 \tablefootmark{d} \\
O$_2$ & 1000 \tablefootmark{c} \\
O$_3$ & 1800 \tablefootmark{e} \\
N$_2$ & 1000 \tablefootmark{c} \\
CH$_4$ & 1300 \tablefootmark{f} \\
NH$_3$ & 1300 \tablefootmark{f} \\
OH & 2820 \tablefootmark{g} \\
H$_2$O & 5640 \tablefootmark{g} \\
H$_2$O$_2$ & 5640 \tablefootmark{g} \\
H$_2$CO & 2050  \tablefootmark{f} \\
CH$_3$OH & 5530 \tablefootmark{c} \\
\hline
\end{tabular}
\label{table_energies}
\tablebib{$^{(a)}$ \citet{Hornekaer2005, AlHalabi2007}; $^{(b)}$ \citet{Tielens1987}; $^{(c)}$ \citet{Collings2004}; $^{(d)}$ \citet{Sandford1990}; $^{(e)}$: \citet{Cuppen2007}; $^{(f)}$ \citet{Garrod2006}; $^{(g)}$ \citet{Speedy1996}}
\end{table}

H$_2$ is, by about four orders of magnitude, the most abundant molecule in molecular clouds. Most particles that accrete onto the surface are, therefore, H$_2$ molecules. At low temperatures, H$_2$ would become the most abundant icy molecule if binding energies were computed relative to a water ice substrate alone. 
However, microscopic models by \citet{Cuppen2007} and \citet{Cuppen2009} have shown that the total binding energy of a adsorbate relative to a substrate is given by the additive energy contribution of the occupied neighbouring sites. Therefore, an H$_2$O:H$_2$ mixture must be considered for computing an effective $E_b$. \citet{Vidali1991} gave an estimate of the binding energy of H on an H$_2$ ice, which is about 45 K at 10 K.
Following \citet{Garrod2011}, we compute the effective binding energy of each species $i$ from the fractional coverage of H$_2$ on the surface, $P$(H$_2$)
\begin{equation} 
E_b (i) = (1-P(\textrm{H}_2)) \cdot E_{b,\textrm{wat}}(i) + P(\textrm{H}_2) \cdot E_{b,\textrm{H}2}(i) \label{EbH} 
\end{equation}
where $E_{b,\textrm{wat}}(i)$ is the binding energy of the species $i$ relative to water ice, the values of selected species are listed in Table \ref{table_energies}. To deduce $E_{b,\textrm{H}2}$ of species other than H, we apply the same scaling factor as for atomic hydrogen ($E_{b,\textrm{H}2}(\textrm{H})/E_{b,\textrm{ASW}}(\textrm{H})$). The increase in the H$_2$ abundance in the mantle tends to decrease the effective binding energy $E_b$($i$) of physisorbed species $i$.

\subsection{Physical model} \label{section: physical}

As described in the introduction, water ice is formed at visual extinctions $A_{V}$ higher than 3 mag \citep[][]{Whittet1988, Whittet2001}, whilst gas phase CO is detected above a visual extinction threshold of 2 mag \citep{Frerking1982} and PDR models show that hydrogen is already mainly molecular at $A_{V} < 1$ mag. Water ice is, therefore, thought to be formed when the gas is already molecular. 

We consider a two-step model. For each set of input parameters, we first compute the abundances of gas phase species assuming the element abundances shown in Table \ref{elem_abu} and considering the gas phase network described in section \ref{gas_phase}. 
{For this, we assume the steady state values, which are reached after $5 \times 10^6$ yr when the density is 10$^3$ cm$^{-3}$ and after $10^6$ yr at $10^4$ cm$^{-3}$. This first step is meant to describe the molecular cloud and the initial pre-collapse phase before the formation of the ice bulk. In practice, we assume that the timescale to reach the chemical equilibrium is shorter than the dynamical timescale for the gas to reach the prestellar core conditions.}
These gas phase abundances are then considered as the initial abundances for the gas-grain modelling. In the second step, we allow gas and grain surface chemistry to evolve whilst physical conditions (density, temperature, visual extinction) remain constant.

\subsection{Multi-parameter approach}

Following our previous works (TCK12a, TCK12b), we consider a multi-parameter approach by considering free physical, chemical, and surface input parameters. This approach allows us to study their influence on the formation and the deuteration of key species. 

The total density of H nuclei in molecular clouds, where interstellar ices are thought to be formed, show typical variations from 10$^3$ cm$^{-3}$ at the edge of clouds to 10$^6$ cm$^{-3}$ in denser prestellar cores. 
Temperatures of the gas and grains also show variations depending on the location inside the cloud. Here, we consider three fixed temperatures, by assuming that the gas and grain surface temperatures are equal, $T_{g} = T_{d}$.
We study the influence of the visual extinction on the formation and the deuteration of ices by considering $A_V$ as a free parameter.
As previously explained in section \ref{gas_phase}, the ortho/para ratio of H$_2$ is considered as constant. 

In section \ref{binding} and in TCK12a and TCK12b, we showed the importance of considering distributions of several grain surface parameters (grain diameter $a_d$, binding energy $E_b$ of light species, diffusion to binding energy ratio $E_d/E_b$). We therefore consider them as free parameters, {the range of values are listed in Table \ref{table_grid}}. 

We keep fixed the following other parameters: \\
- distance between two sites $d_s = 3.1 $ \AA, corresponding to a high-density ASW \citep{Jenniskens1995}; \\
- cosmic-ray ionization rate $\zeta = 3 \times 10^{-17}$ s$^{-1}$; \\
- interstellar radiation field (ISRF) $F_{ISRF} = 1 \times 10^8$ photons cm$^{-2}$ s$^{-1}$; \\
- cosmic ray induced radiation field (CRH2RF) $F_{CRH2RF} = 1 \times 10^4$ photons cm$^{-2}$ s$^{-1}$; \\
- scaling factor in multiples of the local interstellar field $G_0 = 20$, which represents an average value between low interstellar radiation fields seen in molecular clouds, such as B68 \citep[$G_0 = 0.25 -1$,][]{Bergin2002}, and high ISRFs see,n for example, in star-forming molecular clouds in Orion \citep[$G_0 = 10^4$,][]{Giannini2000}.

In the model grid, some of the parameter values are inconsistent with each other and do not necessarily reflect realistic physical models. We use this model grid in order to systematically study the influence of each parameter on the formation and the deuteration of interstellar ices. For this purpose, we compute the mean values and 1-sigma standard deviations of species abundances and deuterations either in the gas phase or grain surfaces, following the method by \citet{Wakelam2010}. 
Table \ref{table_grid} lists all the free parameters and their ranges of values explored in this work.

\begin{table}[htp]
\centering
\caption{List of the input parameters and the values range explored in
  this work. Bold values mark the values adopted in the reference
  models (see text).}
\begin{tabular}{c c}
\hline
\hline
Input parameters & Values  \\
\hline
\textbf{Physical conditions} & \\
$n_{H,ini} $ & $10^{3}$ - ${10^4}$ - $10^{5}$ - $10^{6}$ cm$^{-3}$ \\
$T_g = T_d$ & 10 - {15} - 20 K \\
$A_V$ & 0 - 1 - 2 - 3 - 4 - 5 - 6 - 7 - 8 - 9 - {10} mag \\ 
$\zeta$ & $3 \times 10^{-17}$ s$^{-1}$ \\
ISRF & $1 \times 10^8$ photons cm$^{-2}$ s$^{-1}$ \\
$G_0$ & 20 \\
\textbf{Grain surface parameters} & \\
$a_{d}$ & 0.1 - \textbf{0.2} - 0.3 $\mu$m \\
$E_b($H) & 400 - \textbf{500} - {600} K \\
$E_d/E_b$ & 0.5 - \textbf{0.65} - 0.8 \\
$d_s$ & 3.1 $\AA$ \\
\textbf{Chemical parameters} & \\
H$_2$ o/p ratio & \textbf{$3 \times 10^{-6}$} - $3 \times 10^{-4}$ - $3 \times 10^{-2}$ - 3 \\
\hline
\end{tabular}
\label{table_grid}
\end{table}

\section{Results} \label{results}

In this section, we present the results in two steps. First, we consider the formation of water and other major ice species, and second, we focus on the deuteration. 

Specifically, in section \ref{validation}, we validate our model. 
For this purpose, i) we compare our predicted gas phase abundances as function of the visual extinction with PDR model predictions; ii) we compare the predicted abundances of gaseous and solid water as function of the visual extinction with published observations; iii) we discuss the multilayer formation of interstellar ices for a set of three reference physical conditions. We emphasize that our approach does not pretend to describe the whole evolution of the cloud. The reference models are meant to quantify the influence of specific physical conditions, which are likely to describe different evolutionary stages, on the ice chemistry.

Second, after validating the model, we focus on the water deuteration. In section \ref{ice_ref}, we emphasize the importance of the CO depletion on the deuterium fractionation of the reference models. Then, in section \ref{sec: param_deut}, we perform a multiparameter study that allows us to evaluate the influence of each model parameter on the deuterium fractionation of water. Comparing the model predictions with the observations allows us to constrain a range of values for the chemical and physical parameters.

\subsection{Validation of the model} \label{validation}

\subsubsection{Initial gas phase abundances}

As described in section \ref{section: physical}, we adopt a two-step model. The ice formation phase is followed by considering initial abundances that are computed from steady-state calculations of gas phase chemistry. These initial abundances depend on the density and the temperature, but mainly on the visual extinction $A_V$ (hereafter, $A_V$ means the edge-to-centre visual extinction, half of the observed visual extinction), because photolytic processes play a significant role at low visual extinctions. 
In fact, the abundance of most gas phase species weakly depends on the density and the temperatures (in the range of values considered in this work) but are essentially a function of the visual extinction. Therefore, we compare the gas-phase abundances of key molecules for ice formation (H, D, C, O, CO, O$_2$) between our model and the Meudon PDR code \citep{Lepetit2006} for a total density $n_H = 10^4$ cm$^{-3}$. 

\begin{figure}[htp]
\centering
\includegraphics[width=88mm]{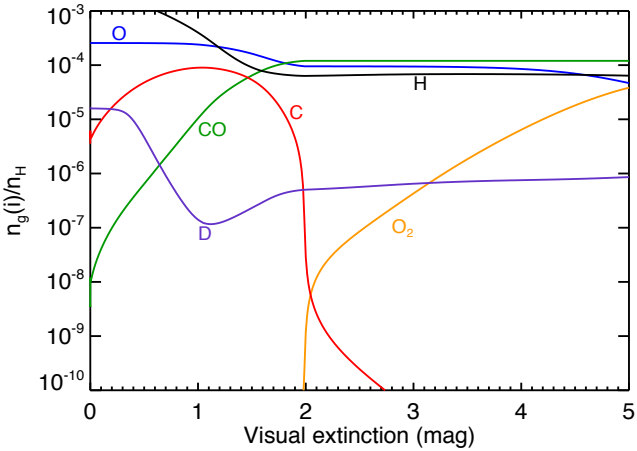}
\caption{Initial abundances of C, O, CO, and O$_2$ as function of the visual extinction computed with the Meudon PDR code at $n_H = 10^4$ cm$^{-3}$, $G_0 = 20$.}
\label{gas_Av}
\end{figure}

Figure \ref{gas_Av} shows the depth-dependent chemical abundances of gas phase H, D, O, C, O$_2$, and CO, for $n_H = 10^4$ cm$^{-3}$ computed by the PDR code in the low visual extinction regime. These abundance profiles are in relative good agreement with the initial abundances computed with our model. The abundances differ by  20\% for most abundant species and within one order of magnitude when species display low abundances (C and O$_2$). This validates the use of the H$_2$, HD, and CO self-shieldings in our code. The other differences are due to the different chemical networks and physical conditions between the two codes (the PDR and ours).
The decrease in UV flux with increasing $A_V$ increases the abundance of molecules such as H$_2$, HD, and CO, and decreases the abundances of H, D, and then C, and O. 
D abundance is governed by the high photodissociation of HD at low $A_V$ and by the formation of H$_3^+$ isotopologues at higher $A_V$, and shows an abundance minimum at $A_V = 1$ mag. 

Molecular oxygen is formed at higher visual extinction, and it reaches its maximal abundance at $A_V > 5$ mag. Our chemical network predicts a high O$_2$/O abundance ratio ($\sim 1 - 1.5$). In fact, the O$_2$ abundance is highly uncertain because it depends on 1) the rate coefficients of some key reactions displaying high uncertainties, and 2) the elemental abundances of carbon and oxygen \citep[see][]{Wakelam2010, Hincelin2011}. 
{Moreover, observations carried out with the \textit{SWASS}, \textit{Odin}, and \textit{Herschel} space telescopes have revealed that O$_2$ is not abundant in molecular clouds \citep[$X($O$_2) \lesssim 10^{-7}$, see][]{Goldsmith2011, Liseau2012}. Modelling the formation of ices with high O$_2$ abundances is not necessarily realistic, and we, therefore, investigate the influence of the gas phase O$_2$/O abundance ratio on the formation of ices in section \ref{part:ref_models}.}

\noindent
\textbf{Concluding remarks.} 
{As shown in Fig. \ref{gas_Av}, the chemistry in the gas phase is known to strongly depend on the visual extinction $A_V$. The steady-state gas phase abundances computed with our model are in good agreement with the results of PDR simulations. The use of a two-step model might overestimate the abundance of gaseous O$_2$ affecting ice formation. Therefore, the influence of the O$_2$/O ratio on ice formation needs to be  investigated. }

\subsubsection{Depth-dependent water abundances}

\begin{figure}[htp]
\centering
\includegraphics[width=88mm]{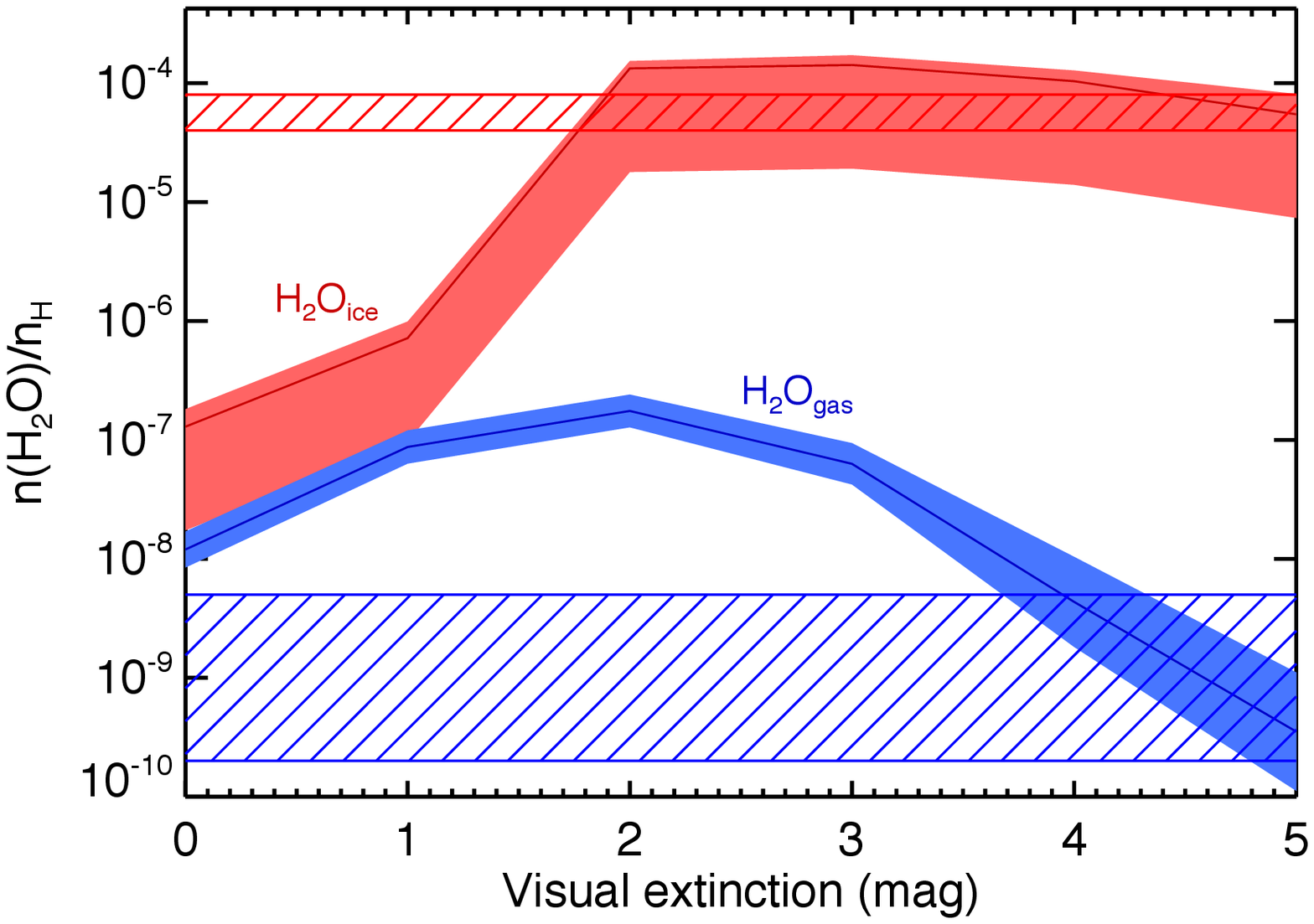}
\caption{Final abundances ($10^7$ yr) of water ice (H$_2$O$_{ice}$, red) and water vapour (H$_2$O$_{gas}$, blue) as function of the visual extinction $A_v$ for typical molecular cloud conditions ($n_H = 10^4$ cm$^{-3}$, $T = 15$ K) and including the variation of all other input parameters. Hatched boxes refer to water ice observed by \citet{Pontoppidan2004} towards the Serpens SMM4 regions and water vapour observed by \citet{Caselli2010} towards the L1544 prestellar core (the upper limit represents water abundance in the external part whilst the lower value represents water abundance in the central region).}
\label{wat_Av}
\end{figure}

Figure \ref{wat_Av} shows the final (at $10^7$ yr) abundances of water ice and vapour as functions of visual extinction $A_V$ for physical condition representative of molecular clouds (e.g. $n_H = 10^4$ cm$^{-3}$, $T_g = T_d = 15$ K). The final abundances can be divided into two zones: 

i) a photon-dominated layer ($A_V < 2$ mag) where gas phase molecules are photodissociated and ices are photodesorbed. 
At very low visual extinctions, interstellar ices are efficiently photodesorbed and show a low abundance of $\sim 10^{-7}$ relative to H nuclei corresponding to less than one monolayer. 
The abundance of water vapour, given by the balance between its formation in gas phase, its photodesorption from interstellar grains, and its photodissociation, reaches steady state values of $10^{-8} - 10^{-7}$ \citep[see also][]{Hollenbach2009}.  

ii) a darker region ($A_V > 2$ mag) where the decrease in UV flux allows the formation of interstellar ices, mainly composed of water. 
At $A_V = 2 -4$, water ice is mainly formed from the accretion of O atoms via the barrierless reaction (\ref{OH_H}). Most of the oxygen reservoir not trapped in CO is easily converted into water ice, reaching abundances up to $2 \times 10^{-4}$. 
The increase in gas-phase O$_2$ initial abundance with $A_V$ slightly decreases the final abundance of water ice. Indeed, water ice is also formed via the accretion of O$_2$ including the formation of hydrogen peroxide. These formation pathways involve reactions possessing significant activation energies and low transmission probabilities (see Table \ref{reacs_eckart}). 
Finally, water formation also depends on the grain surface parameters $E_d/E_b$ and $E_b$(H). A highly porous case (high $E_d/E_b$ ratio combined with a high binding energy of H) strongly decreases the diffusion rate of mobile H species, decreasing the final abundance of water ice by one order of magnitude ($\sim 10^{-5}$).

Gas phase abundance of water is mainly governed by the balance between photodesorption and accretion. The decrease in UV flux with increasing $A_v$ decreases water ice photodesorption and therefore the final abundance of water vapour from $10^{-7}$ at $A_V = 2$ to less than $10^{-9}$ at $A_V = 5$.
We can note that the multilayer nature of grain mantles and the use of a wavelength-dependent absorption of UV photons from ices give similar results to theoretical PDR studies \citep[e.g.][]{Hollenbach2009}.

In spite of the inevitable approximations of our modelling, our predictions are in good agreement with observations of water ice and water vapour, as shown in Fig. \ref{wat_Av}.

\noindent
\textbf{Concluding remarks. }
For typical conditions representative of a molecular cloud, we are able to reproduce the observed $A_V$ threshold ($A_V \sim 1.5$ mag, $A_{V,obs} \sim 3$ mag)  and the high abundance of water ice ($X \sim 10^{-4}$). The low abundance of water vapour observed in molecular clouds is reproduced at higher visual extinctions (4-5 mag).

\subsubsection{Reference models} \label{part:ref_models}

\begin{figure*}[h!btp]
\includegraphics[width=88mm]{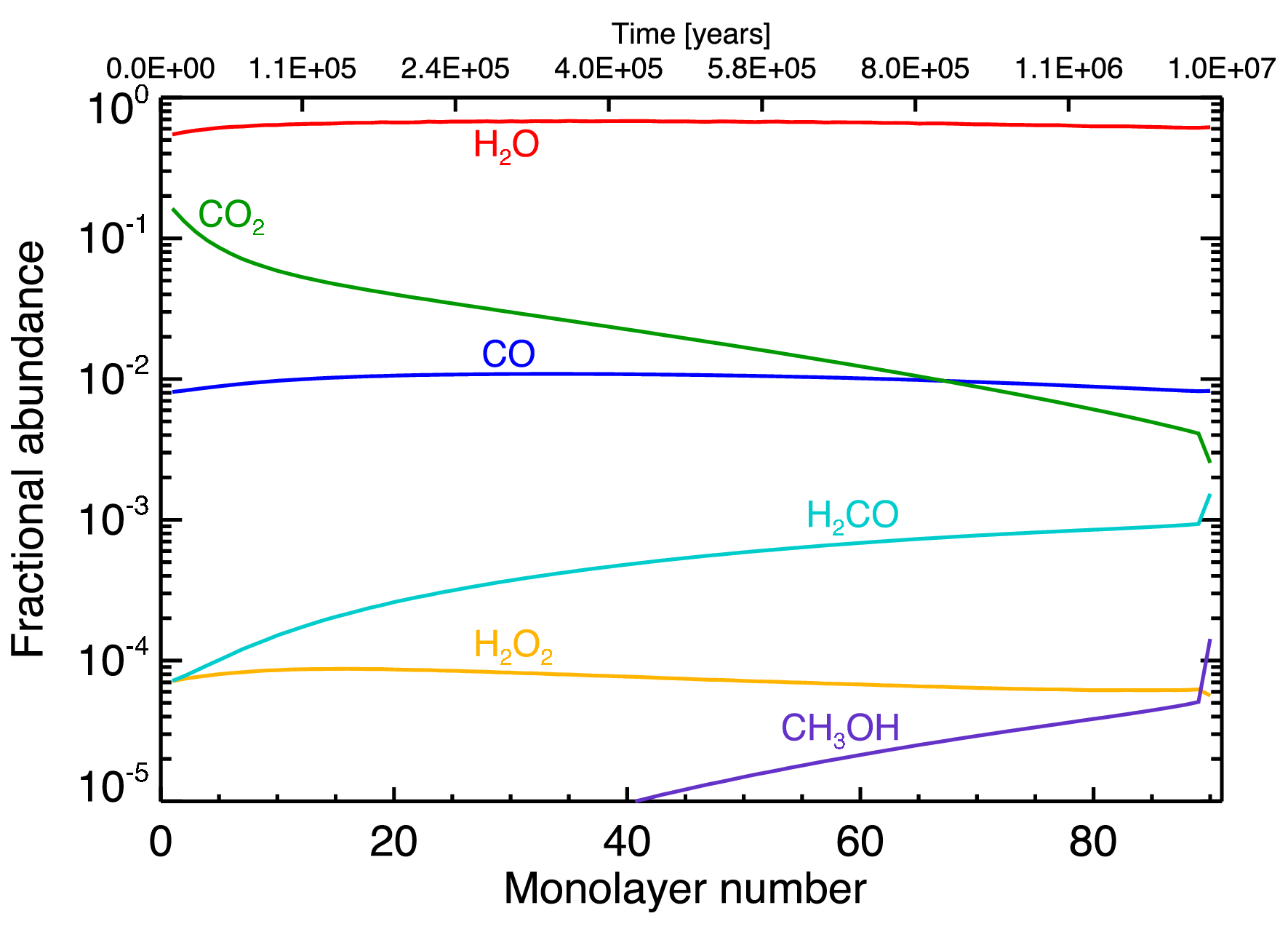} \includegraphics[width=88mm]{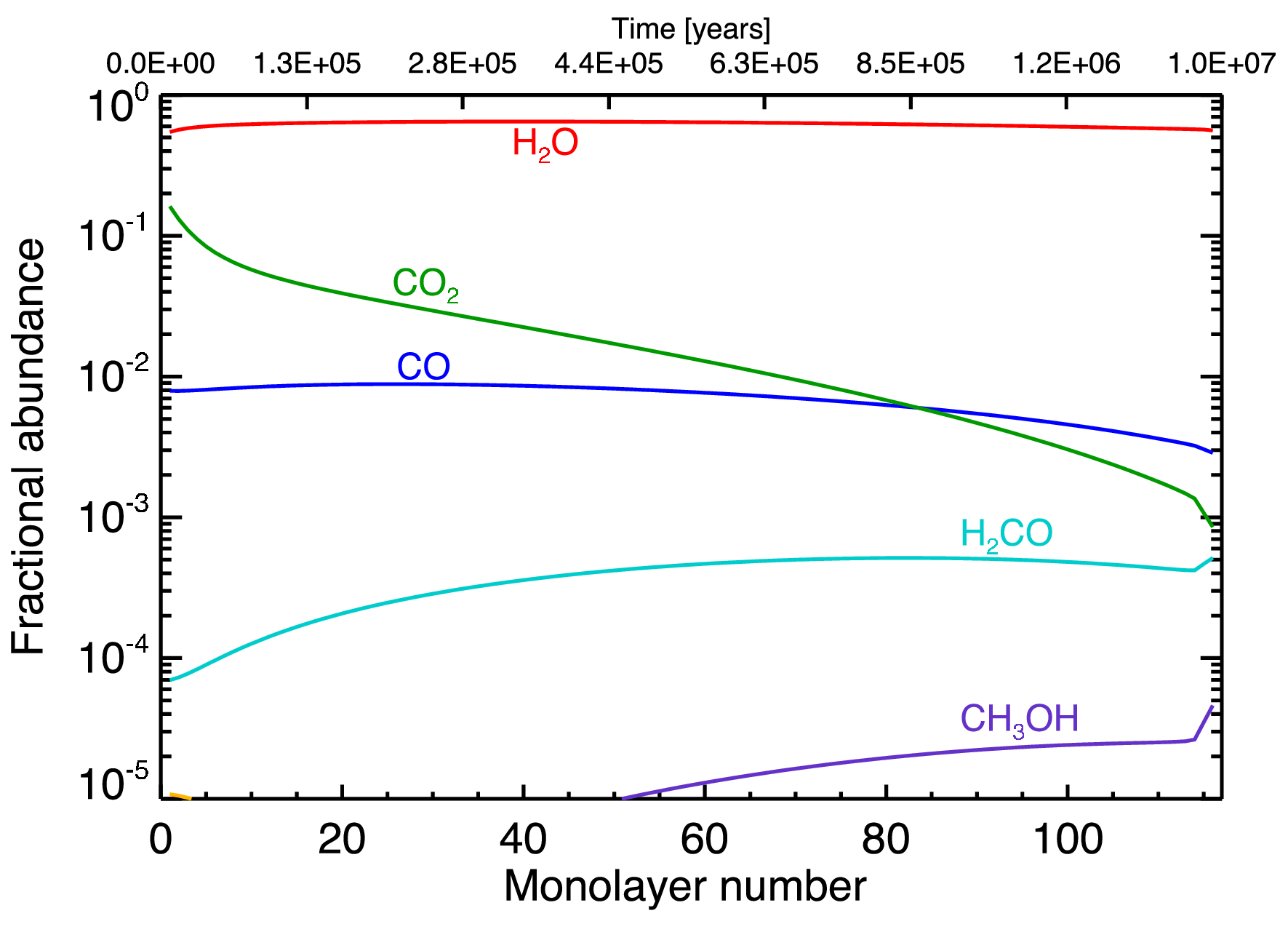} \\ \includegraphics[width=88mm]{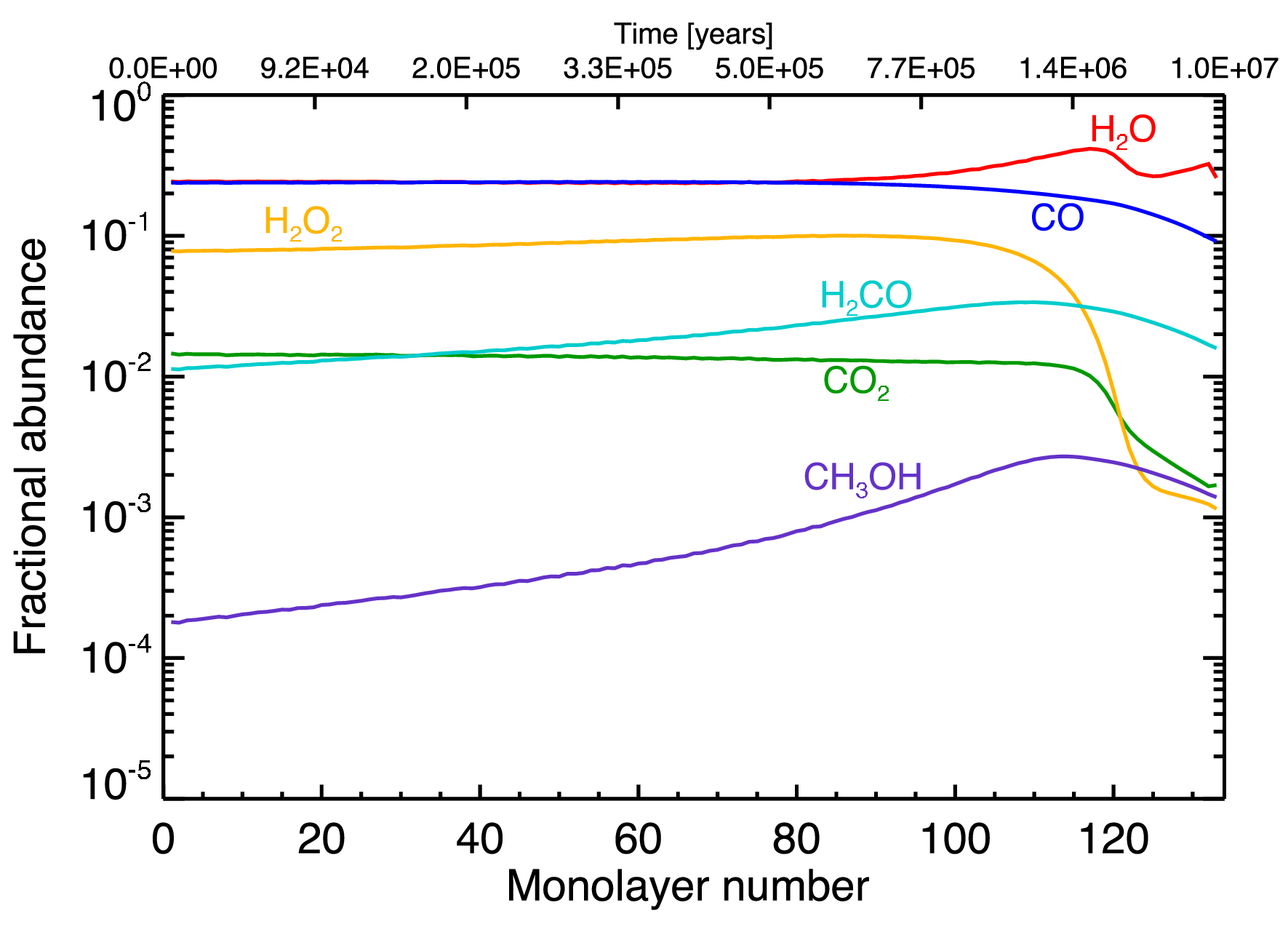} \includegraphics[width=88mm]{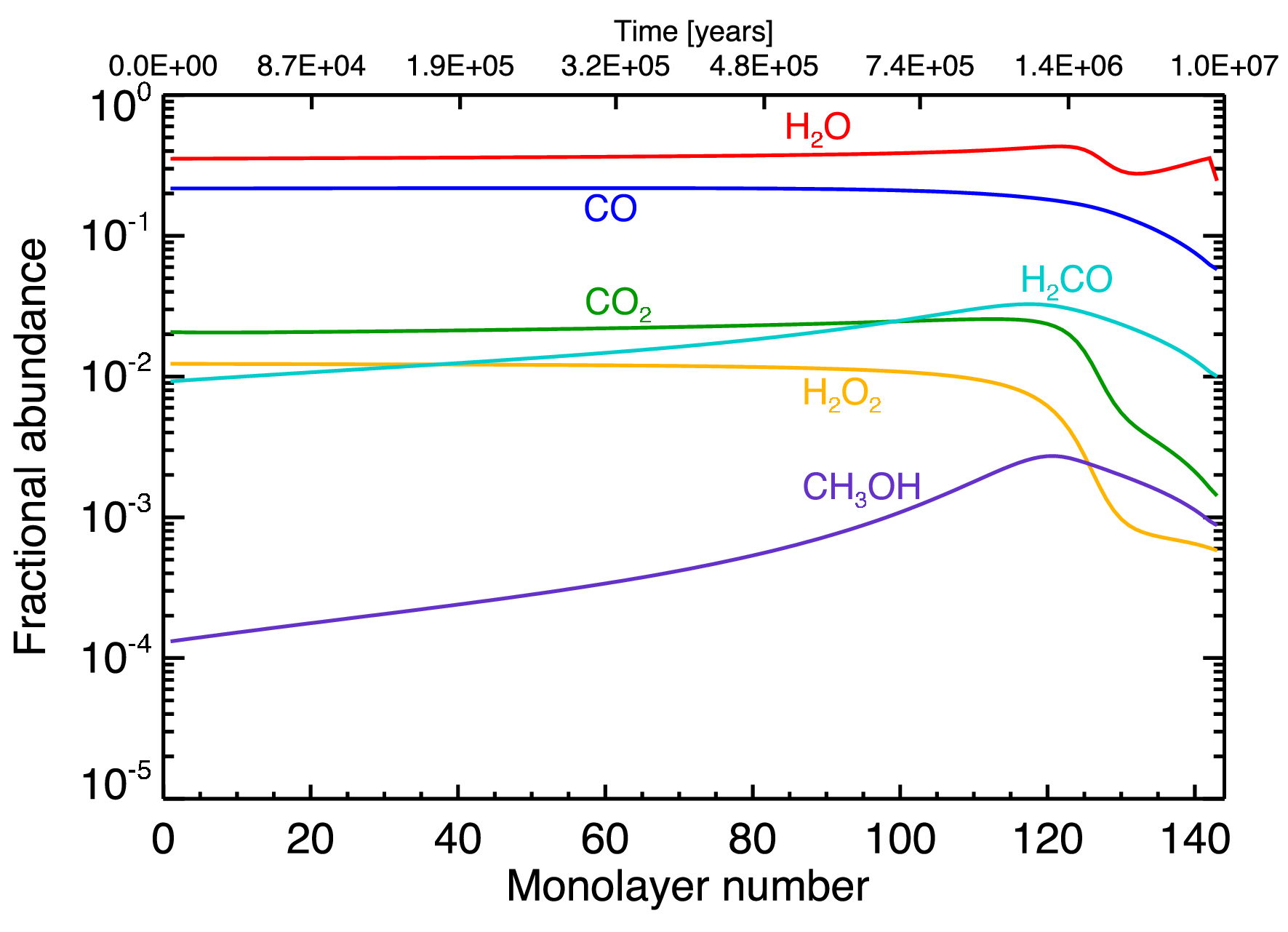} \\ \includegraphics[width=88mm]{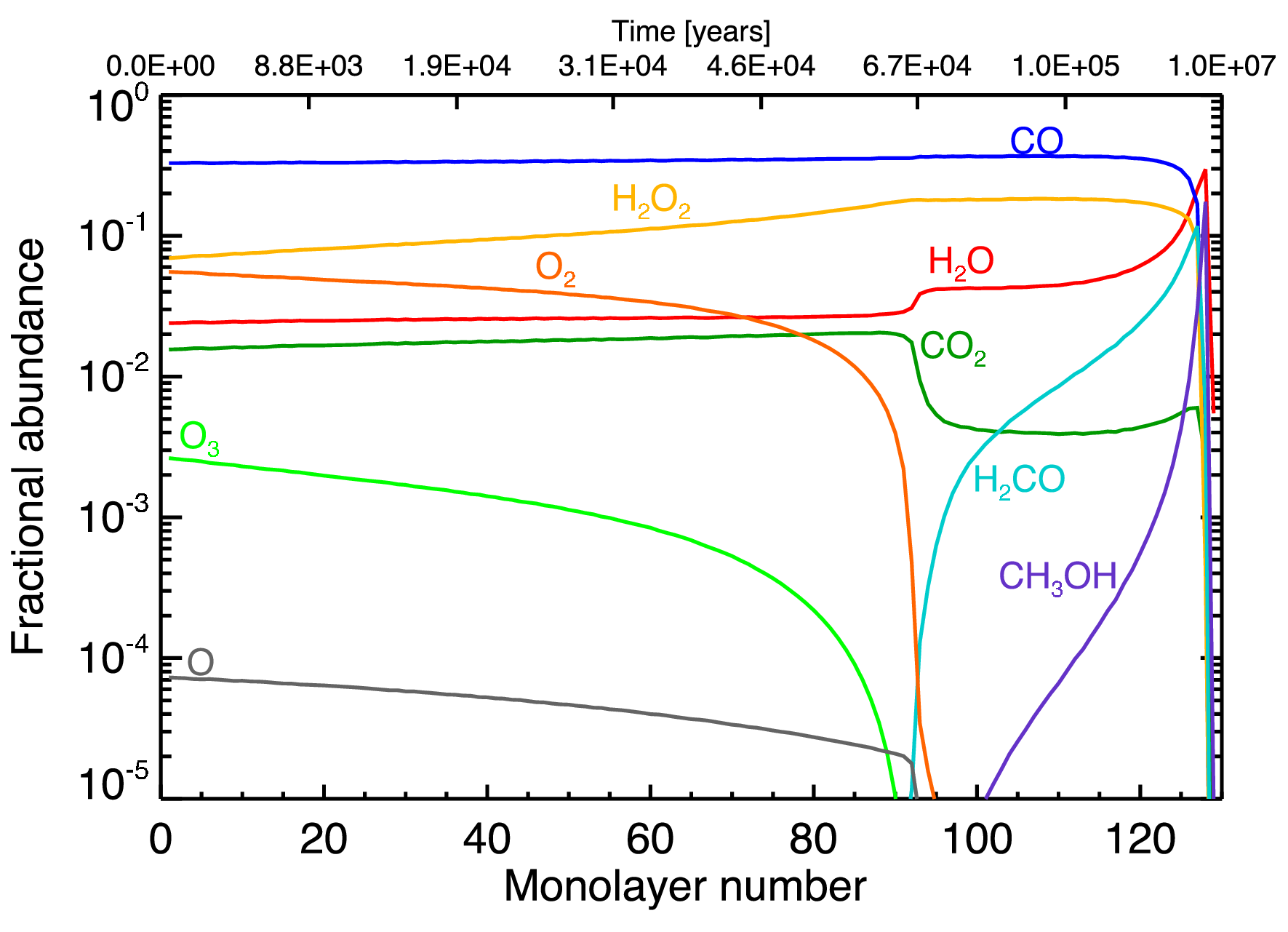} \includegraphics[width=88mm]{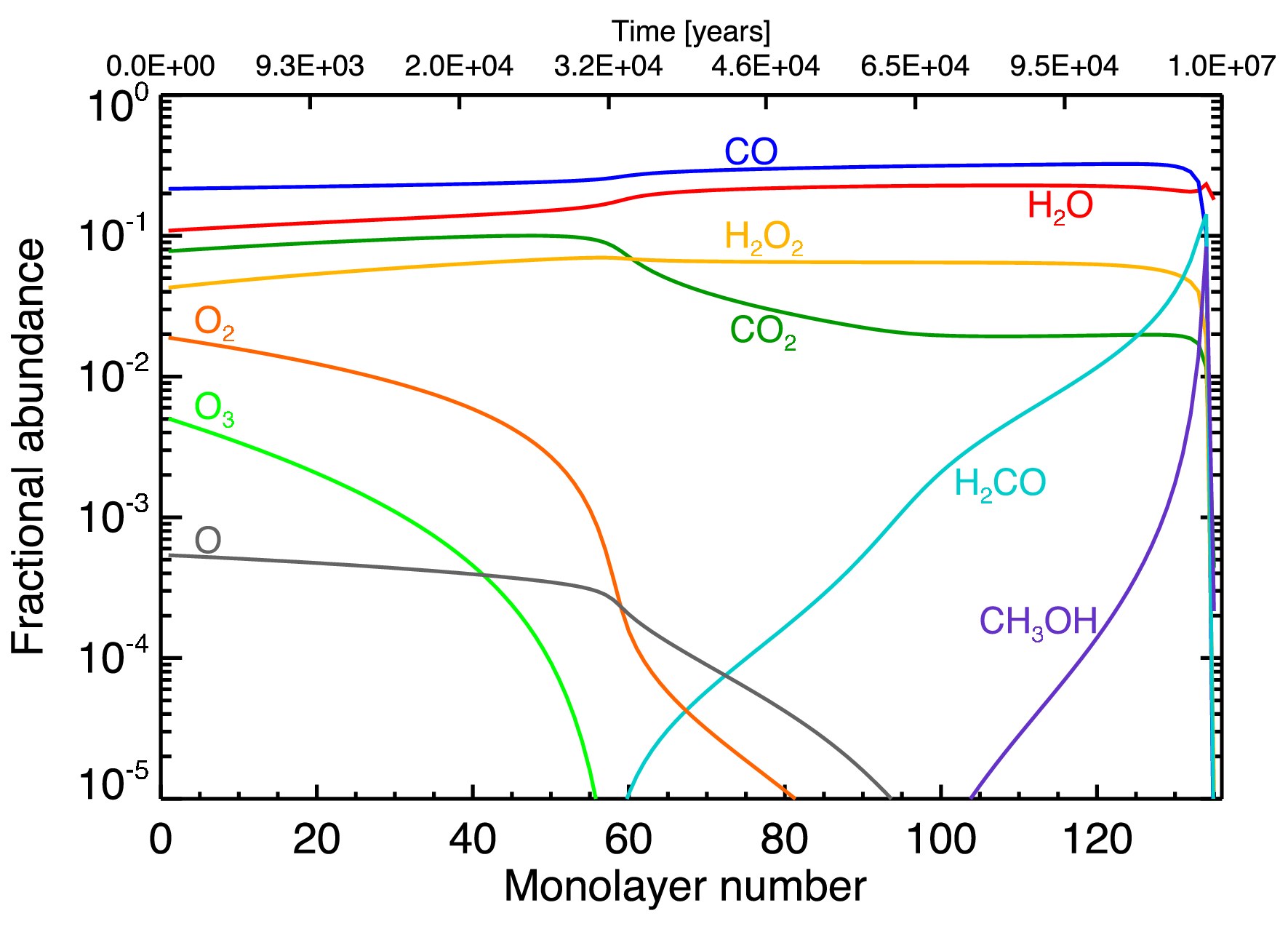} \\
    \caption{{Fractional ice abundances of main species formed from O, O$_2$, and CO for three reference models with ``normal'' gas phase abundances computed with the chemical network presented in section \ref{gas_phase} (left), and for a gas phase [{O$_2$}]/[O] abundance ratio {artificially} decreased by a factor of 10 (right)}: {top)} $n_H = 10^4$ cm$^{-3}$, $T = 15$ K, $A_V = 2$ mag, {middle)} $n_H = 1 \times 10^4$ cm$^{-3}$, $T = 10$ K, $A_V = 4$ mag, {bottom)} $n_H = 10^5$ cm$^{-3}$, $T = 10$ K, $A_V = 10$ mag. Values of other parameters are bold values of Table \ref{table_grid}.}
    \label{ref_models}
\end{figure*}

Very likely, the different ice components and their relative deuteration are the result of a long history where the physical conditions evolve. Therefore, a model aiming at reproducing the whole set of observations should take this evolution into account. However, before embarking on such a complicated modelling, it is worth while and even important to make clear what the characteristics (specifically, solid species and deuteration)are  at each evolutionary step. To this end, we consider three ``reference'' models in the following: 
i) translucent cloud model, forming H$_2$O and CO$_2$ ices; ii) dark cloud model, allowing the formation of CO ice; iii) dark core model, showing high depletion of CO. 
The chemical composition of grain mantles for these reference models are shown in Figure \ref{ref_models} {for the two sets of runs with different [O$_2$]/[O] abundance ratios (see section \ref{gas_phase})}. 

i) \textit{Translucent cloud model} where ices are mainly composed of H$_2$O and CO$_2$: $n_H = 10^4$ cm$^{-3}$, T = 15 K, $A_V = 2$ mag (corresponding to an observed visual extinction of 4 mag). For the conditions considered in this model, grain mantles are mainly composed of water and carbon dioxide. After H$_2$, the most abundant species that accrete onto grain surfaces are H, O, and CO. {For these conditions, the initial O$_2$/O abundance ratio is low ($5 \times 10^{-4}$). The decrease in the O$_2$/O ratio, therefore, does not affect the ice composition significantly. O$_2$ is involved in the formation of O and CO in the gas phase, and its artificial decrease favours O instead of CO and increases the formation of water ice by 25\%, increasing the ice thickness.} The relatively high grain temperature (15 K) allows accreted particles to diffuse efficiently and form H$_2$O and CO$_2$. At this visual extinction, the UV flux irradiating grain surfaces is high, and volatile species, such as atoms or light stable molecules (CO), are efficiently photodesorbed. 

Water is the main ice component because it is formed from the barrierless reactions (\ref{O_H}) and (\ref{OH_H}) whilst its other reaction routes are negligible (less than 0.1 \%). Carbon dioxide is mainly formed from the hydrogenation of the O...CO complex and not by the direct reaction between CO and OH. Indeed, the grain temperature is not high enough to allow a high diffusion of CO and OH, because of their high binding energy. Instead, O atoms that are less attractively bound can meet CO atoms to form the O...CO van der Waals complex. In turn, O...CO readily reacts with H to form a hot HO...CO* complex that forms CO$_2$ + H. 
The accretion of CO and O also allow the formation of formaldehyde and methanol, but only in low abundances, lower than 1 \% compared to solid water {whilst a low fraction of hydrogen peroxide is predicted for a ``normal'' O$_2$/O ratio.} Indeed, they are formed by reactions involving either heavy atoms or significant activation barriers.

The overall abundance of CO$_2$ relative to H$_2$O decreases with time, varying from 30\% at the beginning to less than 1\% at the end. The decrease in the CO$_2$ abundance is due to the increase in H abundance once interstellar ices start to form, because of the high photodesorption. 

Although H$_2$O and CO$_2$ are formed in high abundances, O and CO are not totally depleted on grains. They still show high gas-phase abundances (about $5 \times 10^{-5}$ relative to H nuclei) at the end of the ice formation.
The predicted CO$_2$ abundance relative to water ice is lower than the observations. However, CO$_2$ formation is very sensitive to the grain temperature (governing the diffusion of O atoms) and to the hydrogen abundance (governing the reaction rate between O and H atoms). A higher temperature, higher visual extinction, and/or higher density would, therefore, tend to increase the abundance of CO$_2$ relative to water.

ii) \textit{Dark cloud model} where CO ice starts to form with water, forming a CO:H$_2$O mixture: $n_H = 1 \times 10^4$ cm$^{-3}$, T = 10 K, $A_V = 4$ mag (corresponding to an observed visual extinction of 8 mag), $t = 10^6$ yr. 
The low temperature limits the diffusion of O atoms and heavier species whilst the higher visual extinction allows CO molecules to stay bound on grain surfaces even if they do not react to form CO$_2$. Therefore, the main components of grain mantles are H$_2$O, mainly formed via the reaction (\ref{OH_H}), and CO. 

CO shows an overall abundance relative to water of about 80 \%, which is more than two times higher than observed abundance ratios of ices \citep[10-30 \%,][]{Whittet2007, Oberg2011}. 
This is due to the high O$_2$/O abundance ratio {($\sim 0.5$ throughout the calculation)} that allows the formation of H$_2$O$_2$, via barrierless hydrogenation reactions, instead of water. H$_2$O$_2$ is easily trapped within grain mantles before reacting because the reaction destroying hydrogen peroxide has a low transmission probability (see Table \ref{reacs_eckart}). 
{The decrease in the O$_2$/O ratio by one order of magnitude increases the formation of water, giving a CO/H$_2$O abundance ratio of 30\%, whilst the H$_2$O$_2$ abundance decreases by one order of magnitude. 
Hydrogen peroxide has recently been observed by \citet{Bergman2011}
 towards the $\rho$ Oph A dark cloud, confirming its formation in ices and its subsequent sublimation via non-photolytic processes.} 
The low grain temperature favours the formation of formaldehyde and methanol via CO hydrogenation, compared to the formation of CO$_2$. Formaldehyde shows an unexpected abundance higher than 8 \% compared to water ice. Again, the formation efficiency of formaldehyde and methanol strongly depends on the grain temperature and on the total density. For example, a higher temperature and/or higher density would favour CO$_2$ formation instead of H$_2$CO and CH$_3$OH.

A visual extinction of 4 mag decreases the final gas-phase abundances of O and CO, compared to $A_V = 2$ mag. Indeed, gas-phase CO reaches an abundance of $5 \times 10^{-6}$ whilst O abundance decreases to $10^{-7}$ at $10^7$ yr. The difference in the two abundances is due to the efficient destruction of O atoms forming solid water, whilst most CO molecules do not react and are still able to photodesorb. 

iii) \textit{Dark core model} where most of CO is depleted, allowing the formation of pure CO ice and solid formaldehyde and methanol: $n_H = 10^5$ cm$^{-3}$, T = 10 K, $A_V = 10$ mag (corresponding to an observed $A_V$ of 20 mag). 
The low temperature, the high visual extinction, and the high density allow a significant trapping of CO molecules in the inner part of grain mantles whilst the formation efficiency of CO$_2$ is low owing to the low temperature. {H$_2$CO and CH$_3$OH are formed via hydrogenation reactions that have high activation barriers. They are, therefore, mainly formed in the outer part of the ice when the CO depletion is high, which allows an efficient hydrogenation (see TCK12a). }

At this visual extinction, solid water is less abundant than hydrogen peroxide because the initial gas-phase O$_2$ abundance is higher than the abundance of atomic oxygen {(O$_2$/O between 1 $\sim 3$ throughout the calculation)}. Water formation is mainly formed through the formation of H$_2$O$_2$, which involves reactions having high activation barriers. Hydrogen peroxide is most likely trapped in the bulk before forming water owing to the relatively high density. 
In this case, solid CO reaches an absolute abundance relative to H nuclei of $10^{-4}$, water and CO$_2$ show lower abundances ($10^{-5}$ and $4 \times 10^{-6}$), whilst formaldehyde and methanol show an abundance of $2 \times 10^{-6}$ and $8 \times 10^{-7}$, respectively. 
{Decreasing the O$_2$/O abundance ratio to 0.3 strongly increases the formation efficiency of water and CO$_2$ instead of hydrogen peroxide. For this case, water is almost as abundant as CO. }
Gas-phase abundances of O, O$_2$, and CO decrease with time and show high depletions (with final abundances lower than $10^{-10}$), whereas H abundance remains constant. 
%


\noindent
\textbf{Concluding remarks.}
The chemical composition of ices is very sensitive to the physical conditions and to the initial abundances. Most of the observed ice features are reproduced by the reference models: the water-rich ice seen at low visual extinctions is also composed of abundant CO$_2$, whilst the abundances of solid CO (and  H$_2$CO and CH$_3$OH) gradually increase with the visual extinction and the density. The increase in the O$_2$/O ratio decreases the abundance of water because its formation from O$_2$ involves reactions that have significant activation barriers (see Table \ref{reacs_eckart}). However, since it is likely that O$_2$ abundance remains low in dark clouds, water formation seems to be efficient in a wide range of physical conditions. Therefore, the study of the water deuteration needs to include the variation of several physical parameters.

\subsection{CO depletion and molecular deuteration} \label{ice_ref}

As discussed in TCK12b, the deuteration of solid species strongly depends on the values of the CO depletion and the density at the moment of their formation. 
The initial densities of H and D are roughly constant regardless of the total density. Their abundance relative to H nuclei decreases with increasing $n_{\textrm{H}}$. At low densities, the increase in D abundance is limited by the weak deuterium reservoir, whilst the D/H ratio is able to strongly increase at higher densities.
The gas and grain temperatures, as well as the visual extinction, also affect the evolution of the D/H ratio because they influence the desorption rate of H and D. 
In summary, since all these quantities vary with time, deuteration is not necessarily constant within grain mantles. 

\begin{figure}[htp]
\centering
\includegraphics[width=88mm]{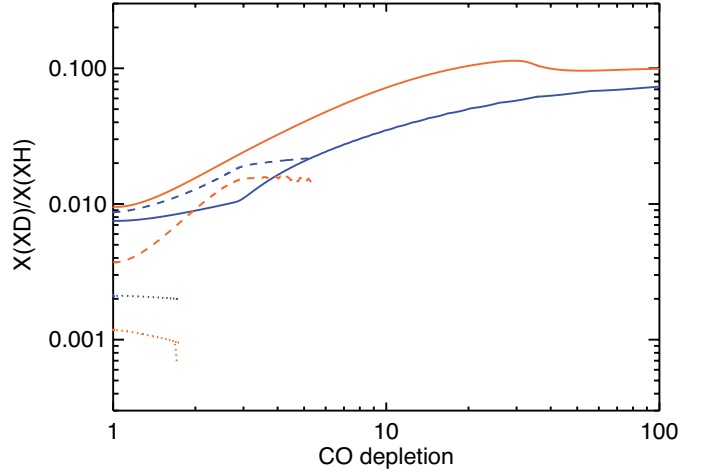}
\caption{Solid HDO/H$_2$O (blue) and gaseous D/H ratio (orange) as function of the CO depletion factor $f_D$(CO) = $n_g$(CO)/$n_{g,ini}$(CO), {which increases with time}, for the three reference models: translucent cloud region (dotted), dark cloud region (dashed), dark core region (solid). 
}
\label{dX_D_COdepl}
\end{figure}

The influence of the CO depletion, the density, the temperatures, and the visual extinction on the water deuteration is shown in Figure \ref{dX_D_COdepl} for the three ``reference'' models described in the previous section.
{At the beginning of the gas-grain calculation, CO has already reached its maximal abundance ($\sim 10^{-4}$, see Fig. \ref{gas_Av}). With time, the CO molecules freeze-out onto grains, decreasing the gas-phase CO abundance and increasing the CO depletion factor.}
The decrease in the temperatures from 15 to 10 K and the increase in the visual extinction from 2 to 4 mag increase the gas-phase D/H ratio from 0.1 \% to 0.4 \% at $f_D$(CO) = 1.  
The increase in total density from $10^4$ to $10^5$ cm$^{-3}$ increases the final D/H ratio from 1 \% to 10 \%.
{The decrease in the gas-phase O$_2$/O abundance ratio does not modify the water deuteration for the two low-density cases. However, it slightly decreases the HDO/H$_2$O ratio for the dense core model by a factor of two because water is more efficiently formed at low CO depletion (see Fig. \ref{ref_models}) when the D/H ratio is low.}

\noindent
\textbf{Concluding remarks.}
As previously noted, water deuteration is largely influenced by the gas D/H atomic ratio. Consequently, the largest deuteration is obtained where the CO depletion and the gas-phase D/H ratio are high, namely in the latest and less efficient phases of water formation, represented by the reference model iii). 
Comparison with the observed deuterium fractionation towards IRAS 16293 shows that a part of water ice should have formed during a dark and/or dense phase. 
In the following section, we investigate the impact of key parameters on water deuteration.

\subsection{Physical/chemical parameters and water deuteration} \label{sec: param_deut}

In this section, we study the influence on ice deuteration of several physical and chemical parameters which play a key role. Each figure presented in this section shows the influence of one (or two) parameter(s) at a time. For each value of the studied parameters, mean value and standard deviation of absolute abundances and deuterations induced by the variation in other input parameters are computed. Comparisons between the evolution of the mean deuteration induced by the variation in the studied parameter and the standard deviation caused by other parameters allow us to deduce the importance of that parameter on water ice deuteration.

The predicted deuteration levels of water ice are compared with the observed HDO/H$_2$O and D$_2$O/H$_2$O ratios towards the hot corino of IRAS 16293 \citep{Coutens2012} to constrain the input parameters that reproduce the observations best. For this purpose, we assume that the entire bulk of interstellar ices desorbed in the hot corino and the observed deuteration reflects the deuteration in ices (see Introduction).

\subsubsection{Influence of the H$_2$ o/p ratio}

\begin{figure}[h!btp] \includegraphics[angle=0,width=\columnwidth,origin=bl]{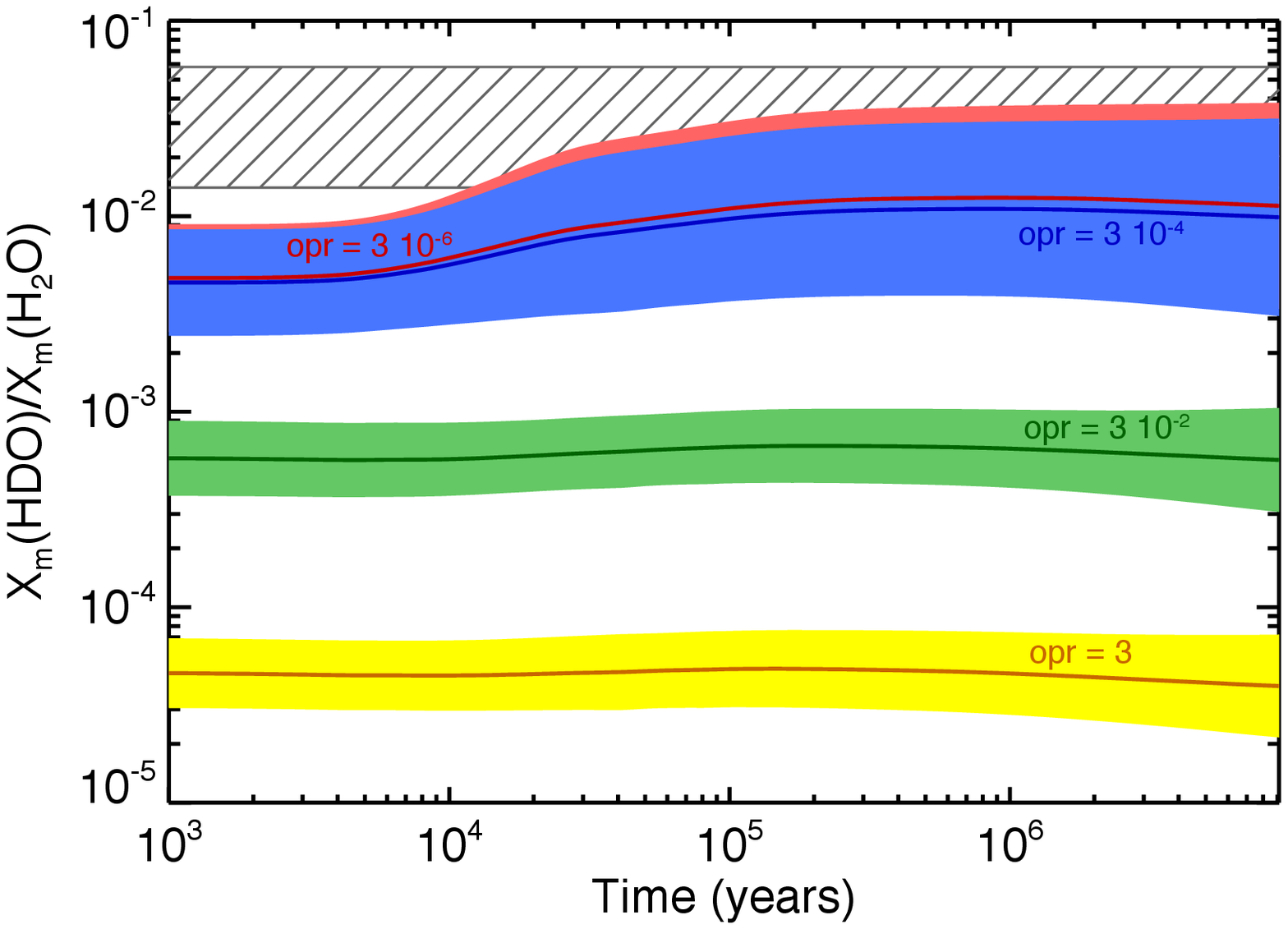}
    \includegraphics[angle=0,width=\columnwidth,origin=bl]{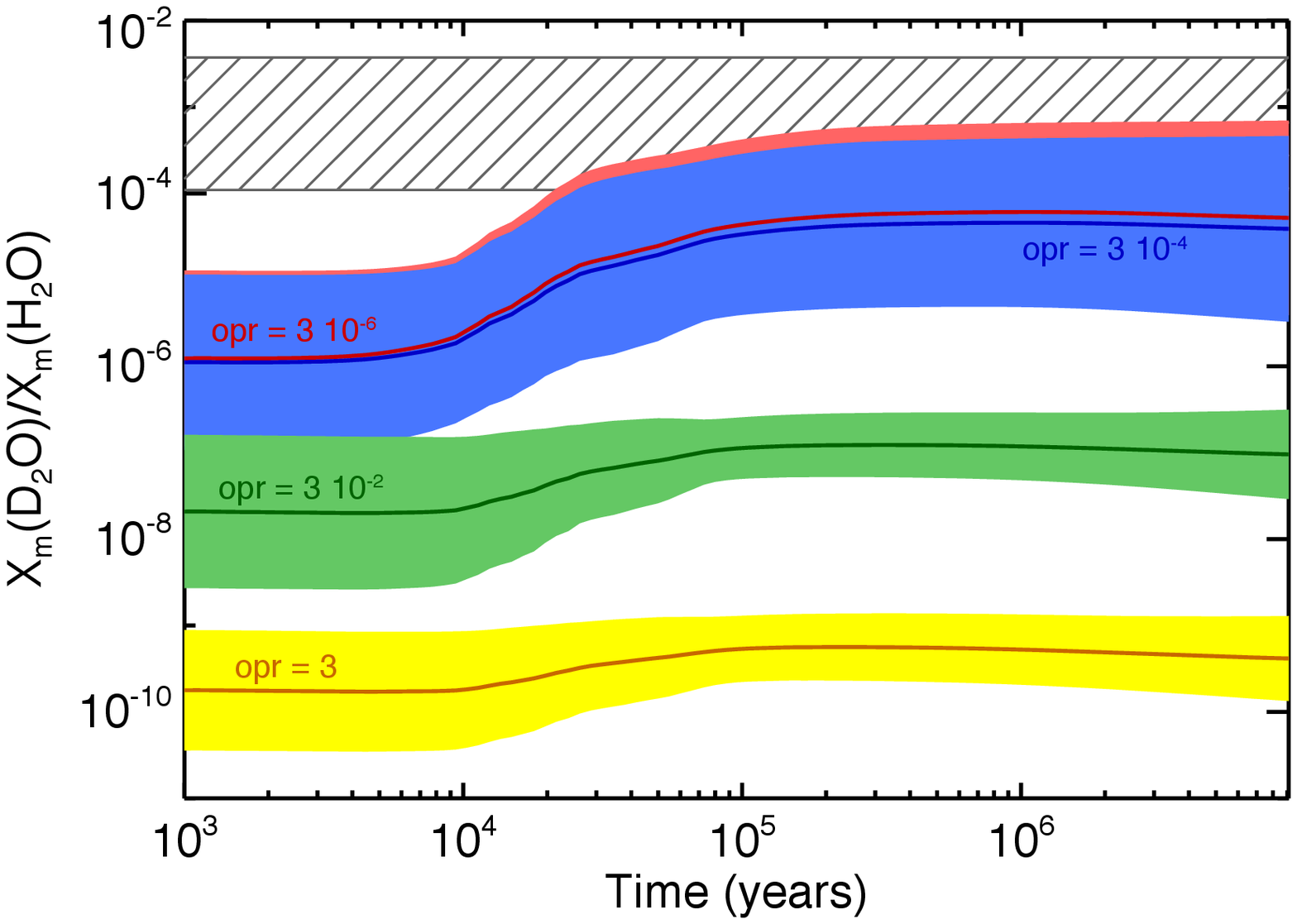}
    \caption{Deuteration of water ice for four values of H$_2$ ortho/para ratios: $3 \times 10^{-6}$ (red), $3 \times 10^{-4}$ (blue), $3 \times 10^{-2}$ (green), and 3 (yellow) including the variation in all other input parameters, except the visual extinction range that is limited to 2 - 10 mag. Hatched boxes refer to water deuteration observed by \citet{Coutens2012} towards IRAS 16293.}
    \label{deut_opr}
\end{figure}

As described in section 2.2, the H$_2$ opr influences the deuteration of gas-phase species (including atomic D) \citep{Walmsley2004, Flower2006, Pagani2009}, thereby affecting the deuteration of solid water.
Figure \ref{deut_opr} shows the HDO/H$_2$O and D$_2$O/H$_2$O ratios in grain mantles as a function of time for four values of H$_2$ opr, including the variation in all other parameters, except the visual extinction range, which is limited to 2 - 10 mag (where most of water is believed to form, see Fig. \ref{wat_Av}). As anticipated, the H$_2$ opr governs the deuteration of water ice via its influence on the abundance of H$_3^+$ isotopologues and atomic D, for values higher than about $3 \times 10^{-4}$.
An increase in H$_2$ opr between $3 \times 10^{-4}$ and 3 decreases the HDO/H$_2$O and D$_2$O/H$_2$ ratios  by 2.5 orders and 5.5 orders of magnitude, respectively. 
Furthermore, the ortho/para ratio of H$_2$ is a key, even the most important, parameter for water deuteration. 
The decrease in the deuteration due to the increase in the H$_2$ opr is much greater than the standard deviation induced by the variation in other parameters.

Comparisons with the observations by \citet{Coutens2012} clearly suggest that a low opr of H$_2$, lower than $3 \times 10^{-4}$, is needed to reproduce the observed deuterium fractionation. 

\subsubsection{Influence of the total density}

\begin{figure*}[h!btp]
  \includegraphics[width=88mm]{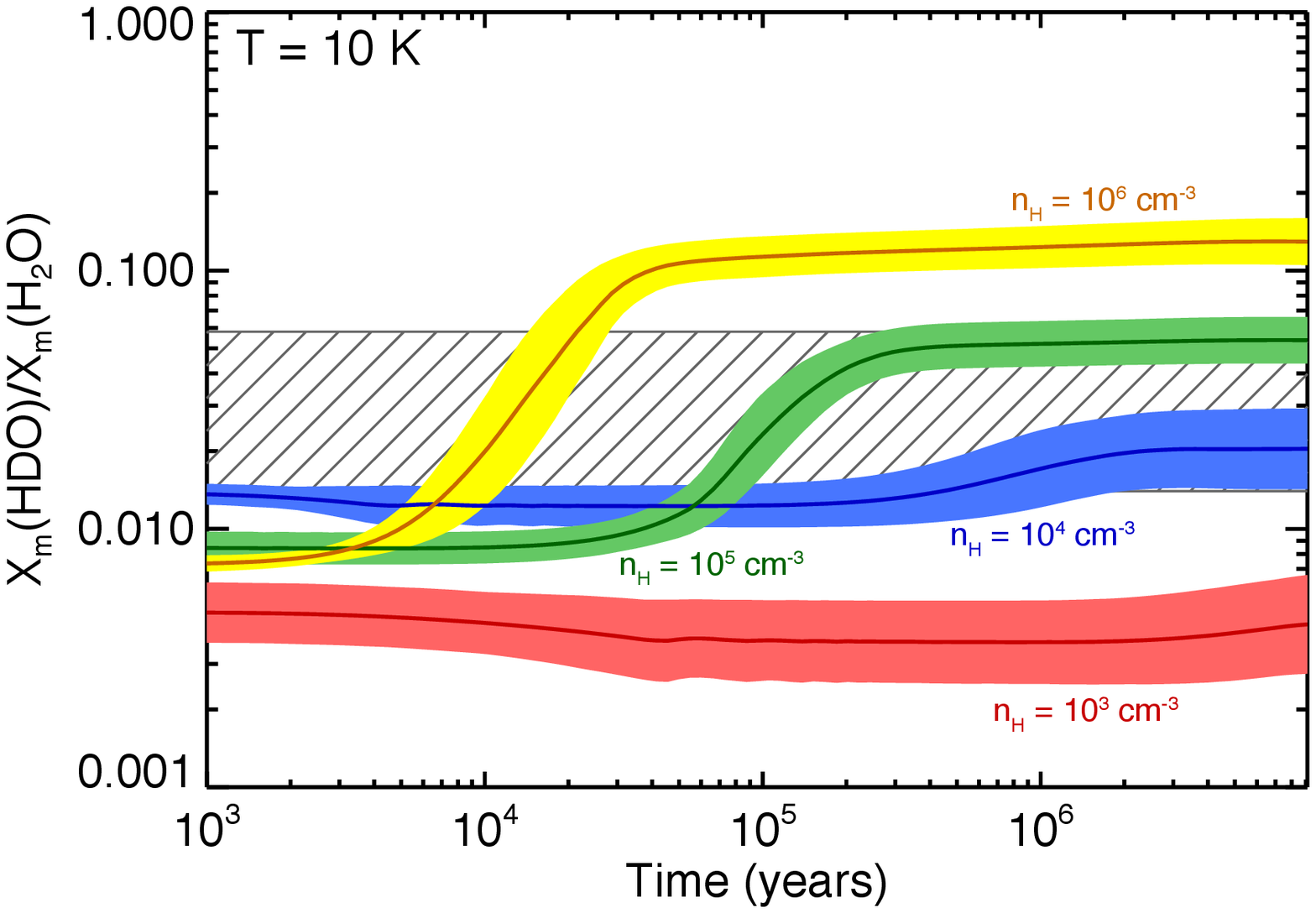}
  \includegraphics[width=88mm]{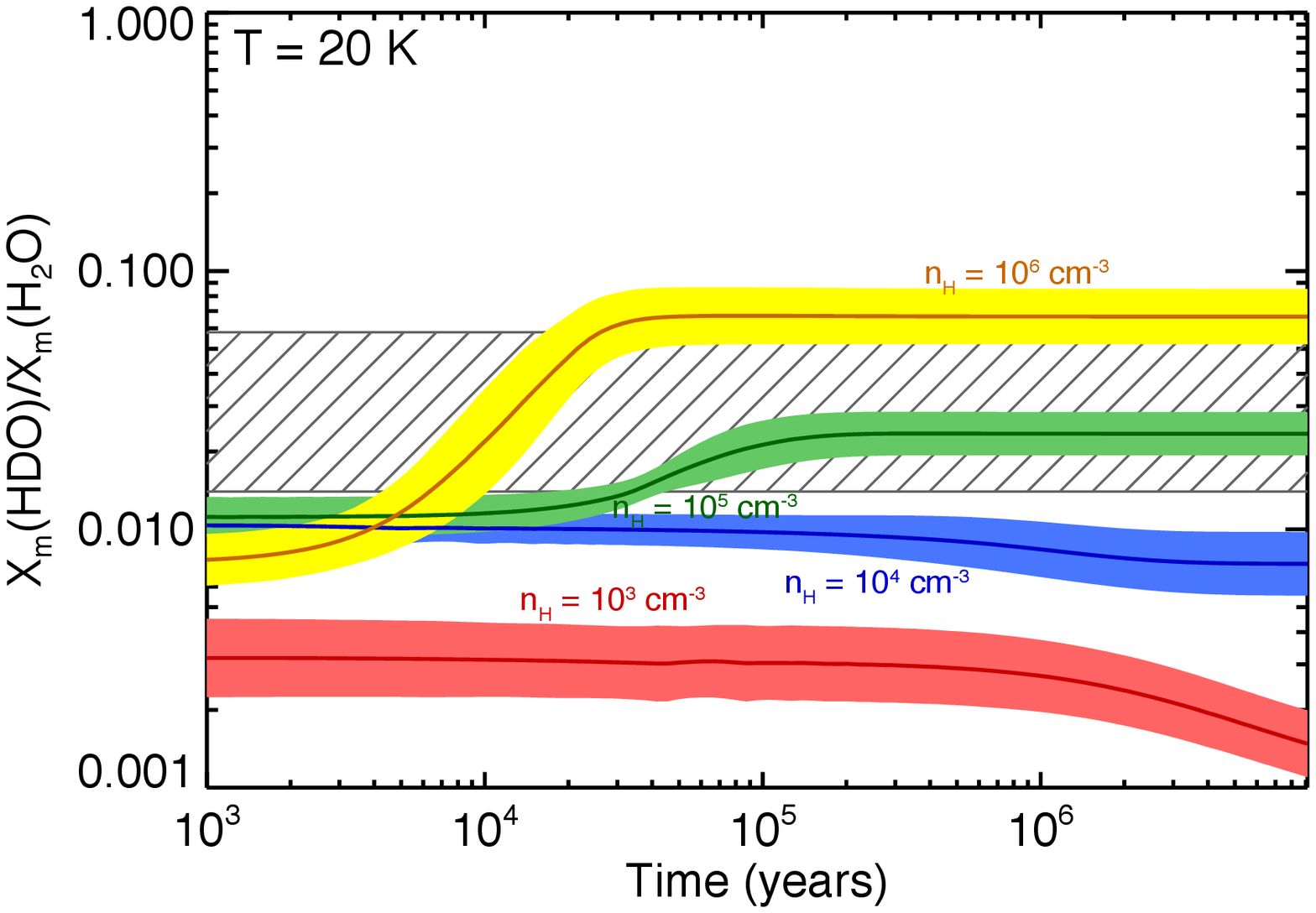} \\
  \includegraphics[width=88mm]{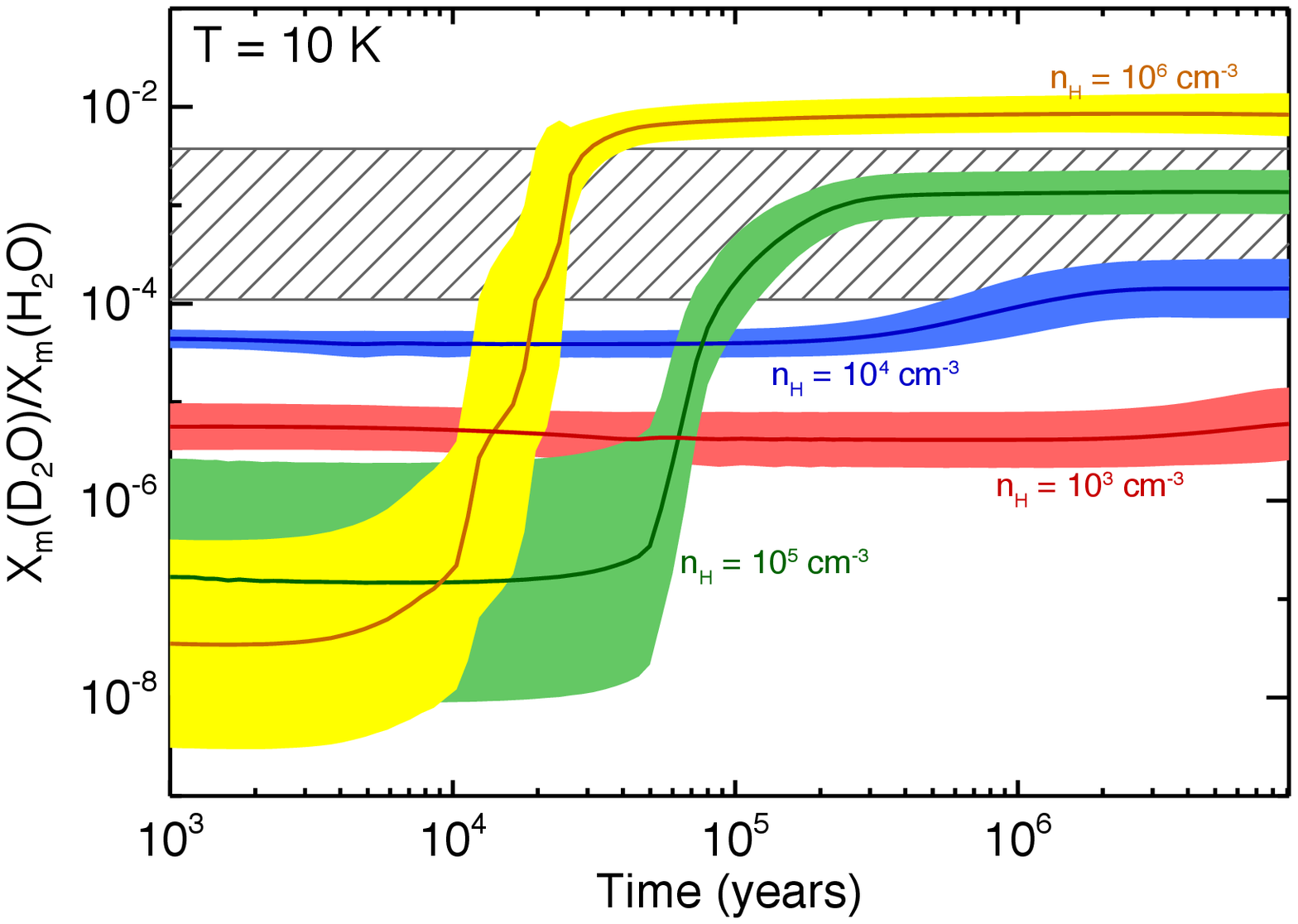}
  \includegraphics[width=88mm]{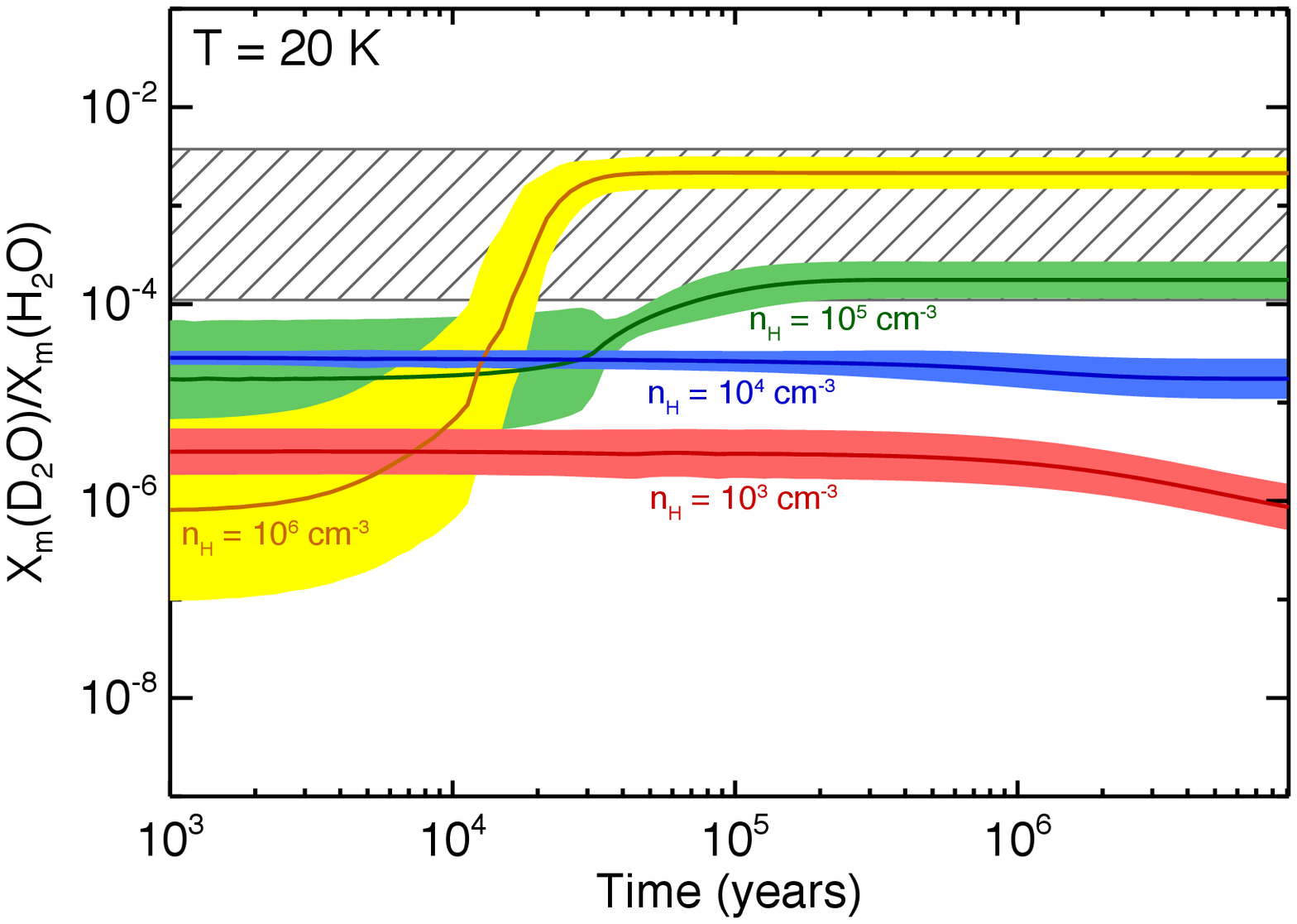}
    \caption{Deuteration of water ice for four values of total density $10^3$ (red), $10^4$ (blue), $10^5$ (green), $10^6$ (yellow) cm$^{-3}$, and two temperatures: 10 K (left panels), 20 K (right panels), using an H$_2$ opr of $3 \times 10^{-6}$, a visual extinction of 10 mag, including the variation of all other input parameters. Hatched boxes refer to water deuteration observed by \citet{Coutens2012} towards IRAS 16293.}
    \label{deut_nH}
\end{figure*}

As discussed in section \ref{ice_ref}, the gas-phase D/H ratio is a function of the total density $n_{\textrm{H}}$ because high densities allow the gas-phase D/H ratio to increase with the CO depletion. Figure \ref{deut_nH} shows the deuteration of HDO and D$_2$O for the four considered densities, using an H$_2$ opr of $3 \times 10^{-6}$, two temperatures (10 and 20 K), and a high visual extinction (10 mag), including the variation in grain surface parameters. 

At low densities ($n_H \leq 10^4$ cm$^{-3}$), water deuteration is constant and low with time because a significant part of HD is already trapped in atomic D, before the CO depletion. 
At higher densities ($n_H > 10^5$ cm$^{-3}$), the efficient increase in the gaseous D/H ratio allows an increase in the deuteration of water up to $10$\%. In these cases, HDO is mostly located in the outer part of grain mantles, and D$_2$O is only located in the outermost layers, whilst H$_2$O is present throughout the mantle bulk. For this reason, the HDO/H$_2$O is limited and cannot reach the final gas-phase D/H ratio (up to 50 \% at $n_H = 10^6$ cm$^{-3}$).

In summary, the total density plays a key role in the deuteration of water. Indeed, the variation in the total density $n_{\textrm{H}}$ between $10^3$ and $10^6$ cm$^{-3}$ influences the HDO/H$_2$O and D$_2$/H$_2$O ratios by 2 and 3.5 orders of magnitude, respectively. 

{The two observed deuteration ratios can be predicted with total densities $n_{\textrm{H}}$ between $10^4$ and 10$^5$ cm$^{-3}$ at 10 K and between $10^5$ and 10$^6$ cm$^{-3}$ at 20 K, regardless of other grain surface parameters. 
In the following section, we study the case $n_H = 10^4$ cm$^{-3}$ in more detail to investigate the effect of the temperature and the visual extinction on water deuteration.}

\subsubsection{Influence of the temperature and visual extinction}

Temperatures either in the gas-phase and on grain surfaces also influence the deuteration of water. First, an increase in the gas-phase temperature enhances the reactivity of endothermic reactions, which can hydrogenate back H$_3^+$ isotopologues, decreasing the abundance of gaseous D atoms. Second, an increase in the grain temperature severely increases the desorption rate of volatile species, such as atomic H or D. The abundance of gaseous D is limited by the low deuterium reservoir, whereas the abundance of atomic hydrogen can increase up to two orders of magnitude, decreasing the gaseous D/H ratio. Water is mainly produced via reactions involving atomic H and D, increasing gas and dust temperatures, hence, decreasing the water deuteration on grain surfaces. 

\begin{figure}[h!btp]
  \includegraphics[angle=0,width=\columnwidth,origin=bl]{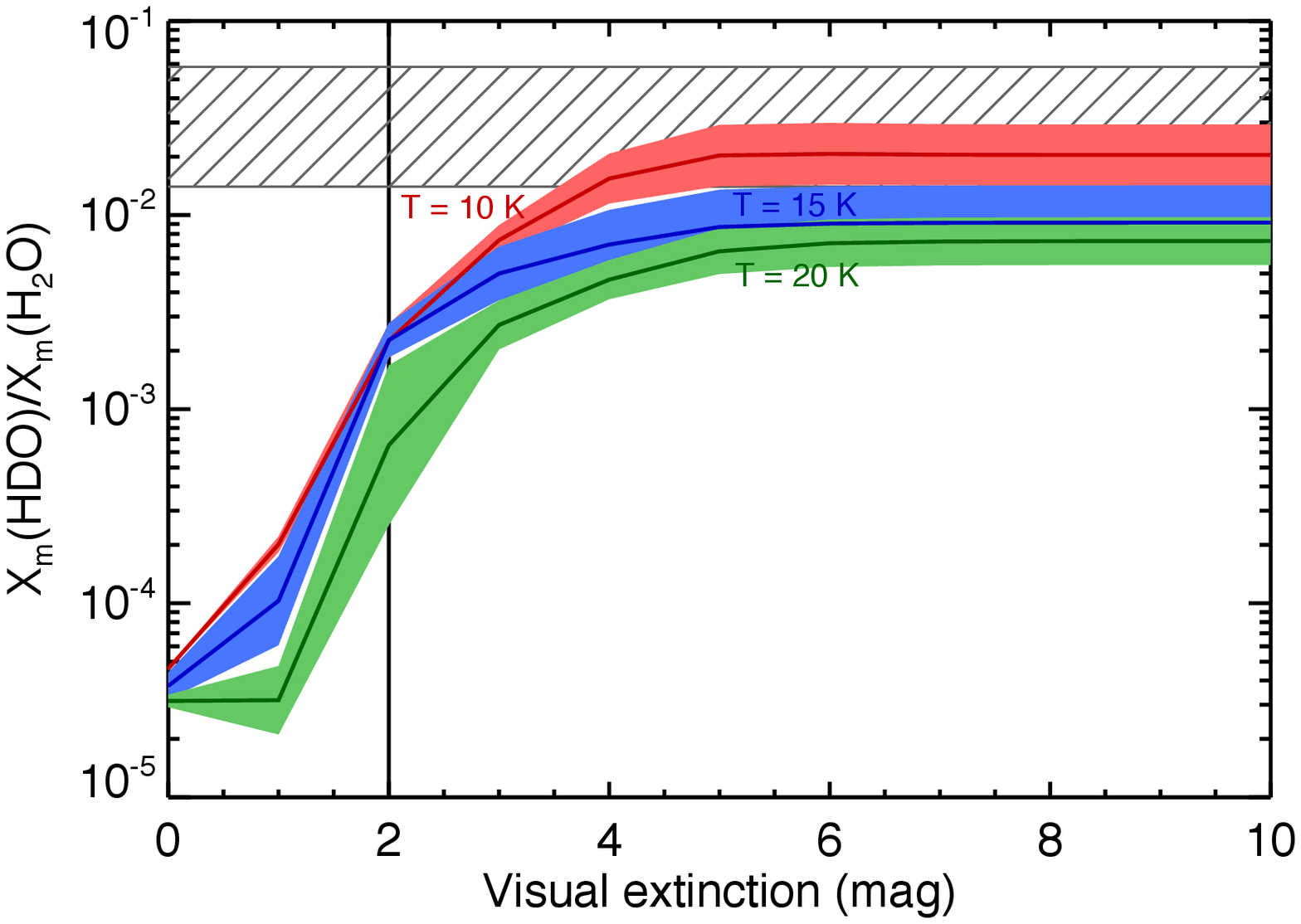}
    \includegraphics[angle=0,width=\columnwidth,origin=bl]{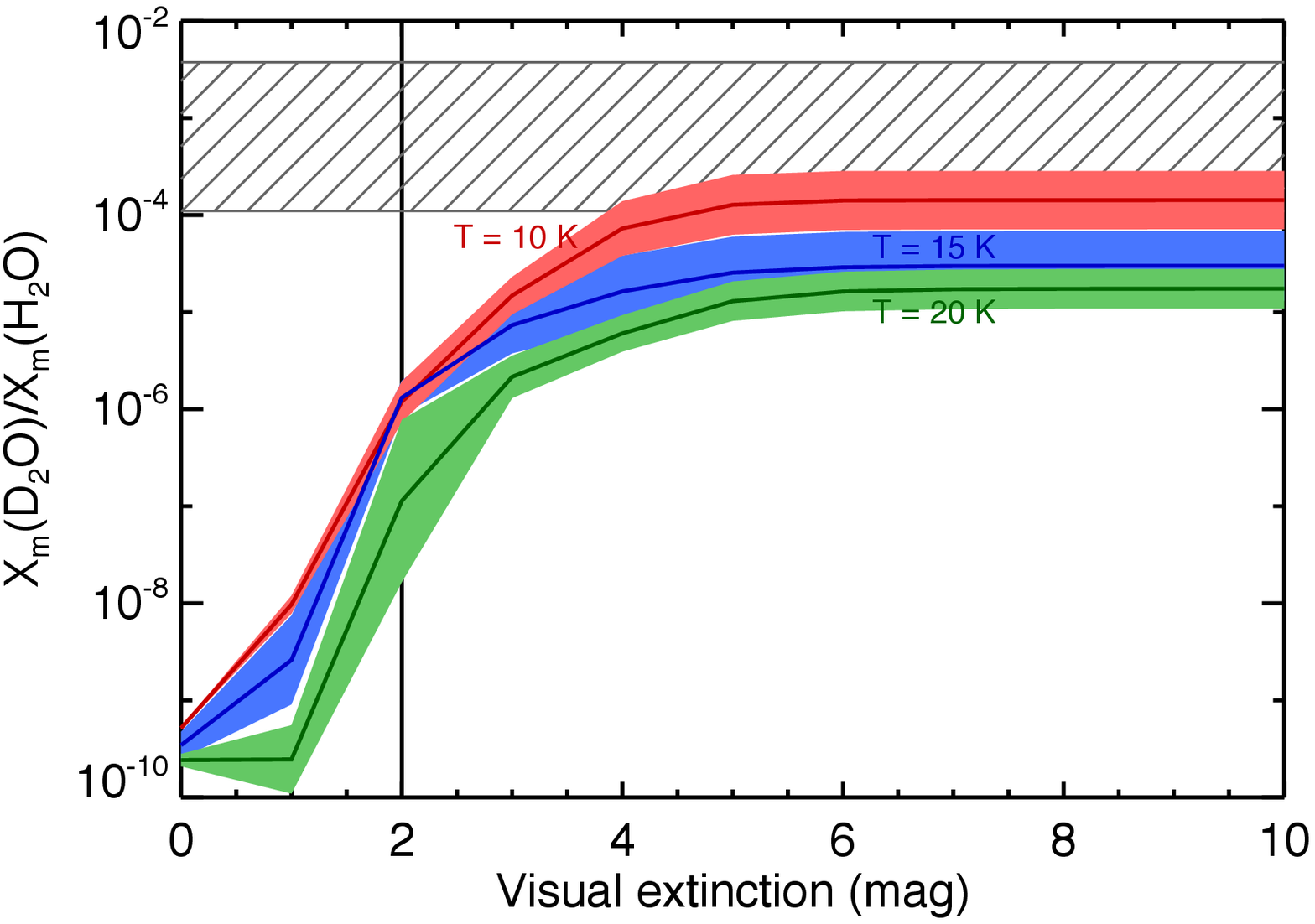}
    \caption{Final deuteration of water ice (at {$t > 10^6$ yr: representative of a typical molecular cloud age}) as function of visual extinction for three values of temperatures: $10$ (red), $15$ (blue), $20$ (green) K, using a total density of $10^4$ cm$^{-3}$, an H$_2$ opr of $3 \times 10^{-6}$ and including the variation of all other input parameters. The vertical line refers to the $A_V$ threshold of water ice observed by \citet{Whittet1988}. Hatched boxes refer to water deuteration observed by \citet{Coutens2012} towards IRAS 16293.}
    \label{deut_T}
\end{figure}

Figure \ref{deut_T} shows this effect by presenting the final HDO/H$_2$O and D$_2$O/H$_2$O ratios (at $10^7$ yr) with the visual extinction, for the three temperatures, considering an H$_2$ opr of $3 \times 10^{-6}$, a total density of $10^4$ cm$^{-3}$, and including the variation in grain surface parameters. 
The increase in the (gas and grain) temperatures between 10 and 20 K slightly decreases the water deuteration approximately by a factor of 3 for HDO and of 10 for D$_2$O at visual extinctions higher than 2 mag. The evolution of deuteration is the same order of magnitude as the standard deviations induced by the variation of other parameters. 

The H and CO abundances in the gas-phase increase with decreasing $A_V$ because of the increase in the photodesorption rate, decreasing the gaseous D/H ratio. 
Visual extinction, therefore, influences water deuteration at low visual extinctions where water starts to form (2mag $<A_V<5$ mag). As can be seen in Fig. \ref{deut_T}, HDO/H$_2$O and D$_2$O/H$_2$O ratios increase by one and two orders of magnitude respectively, between $A_V = 2$ and 5 mag, .

{Comparison with observations shows that observed HDO and D$_2$O fractionations are reproduced for a low temperature of 10 K and for visual extinctions $A_V$ higher than 4 mag ($A_{V,obs} = 8$ mag), implying that deuterated water needs to be formed in dark regions if $n_H = 10^4$ cm$^{-3}$.}

\subsubsection{Influence of grain surface parameters}

The grain surface parameters (${a_d, E_d/E_b, E_b}$(H)) can also influence the formation of interstellar ices. As discussed in TCK12a, the absolute abundance of the main ice constituents formed on grain surfaces (water, formaldehyde, methanol for instance) decrease with ${E_d/E_b}$ whilst the grain size does not affect the overall abundance (but only the ice thickness). Moreover, it is also seen that the absolute abundance of water slightly decreases with the binding energy of volatile species.

The water deuteration is slightly affected by the variation of the grain surface parameters. As shown in Fig. \ref{deut_T}, the standard deviations given by the variation of the grain surface parameters remain lower than the evolution of the water deuteration lead by the variation in the temperature and the visual extinction. 
The water deuteration slightly decreases with the grain size due to the decrease in the CO depletion efficiency (since the accretion rate is inversely proportional to the grain size), limiting the increase in the deuterium fractionation.
On the other hand, the water deuteration slightly increases with ${E_d/E_b}$ and ${E_b}$(H) because the formation efficiency of H$_2$O decreases with the diffusion energy of volatile species, increasing the overall fractionation.

\subsubsection{Concluding remarks}

{Comparing the average value of the deuterium fractionation with the standard deviation induced by the variation of other parameters allows us to claim that the H$_2$ opr is the most important parameter for water deuteration, followed by the total density. The visual extinction and the temperature also influence water deuteration but more weakly. }

The comparison of our model predictions with observations allows us to constrain some parameters involved in the formation of deuterated water seen towards IRAS 16293. The observations are reproduced for \\
- a H$_2$ opr lower than $3 \times 10^{-4}$; \\
- a density lower than $10^5$ cm$^{-3}$ if $T = 10$ K and $A_V > 4$ mag; \\
- a higher density $n_{\textrm{H}}$ between $\sim 5 \times 10^4$ cm$^{-3}$ and $10^6$ cm$^{-3}$ if $T =20$ K.

\section{Comparisons with previous models} \label{comp_models}

In this section, we compare our model predictions for the gas-phase deuteration and for water ice deuteration with previous models. 
The atomic D/H ratio in the gas-phase depends on the deuteration of H$_3^+$. We compare our predictions from steady state models with those by \citet{Flower2006}, specifically their Fig. 5. We obtain the same influence of the H$_2$ opr on the deuteration of H$_3^+$, with similar fractionation values. 

To our knowledge, the only comprehensive theoretical study focused on the deuterated water formation on grain surfaces was performed by \citet{Cazaux2011}, even if other works also included the deuteration of water ice in their astrochemical model \citep[i.e.][]{Tielens1983, Caselli2002, Stantcheva2003, Bell2011}. \citet{Cazaux2011} considered a static stage followed by a free-fall collapse phase. However, in their model, the deuterated water is practically formed only during the first static phase. Thus, it is worth comparing their predictions with ours.

The main difference is that they found that the water deuteration is highly temperature-dependent and lower than our predictions at low ($< 15$ K) temperatures \citep[Fig. \ref{deut_nH} of this work versus Fig. 3.c of][]{Cazaux2011}. This is due to a different approach in the reaction probability computation and the use of the endothermic O+H$_2$ reaction in the chemical network. \citet{Cazaux2011} assumed a competition between the reaction and the diffusion. In their model, if the diffusion timescale of the reactants is larger than the transmission timescale of the reaction, the reaction occurs regardless of its activation barrier. Thus, at low temperatures, the mobility is low and the reaction always occurs. In contrast, in our work, the reaction rate is given by the product of the collision rate of the two reactants and the reaction probability $P_r$ (see Sec. 2.1 and 2.4), giving a much smoother dependence on the temperature. 
As a consequence,  in \citet{Cazaux2011}, H$_2$O is formed, at low temperatures, via reactions involving H$_2$ because of its large abundance, whilst HDO is formed via reactions involving atomic D. The solid HDO/H$_2$O ratio, therefore, scales with the gaseous [D]/[H$_2$] ratio ($\sim 10^{-5}$). In our work, water is always mostly formed via the O+H and OH+H reactions, and the deuteration scales with the [D]/[H] ratio (up to $10^{-1}$).
At higher temperatures ($> 15$ K), both models agree since water is formed via the same reactions, namely O+H and OH+H (and their deuterated counterparts).
This comparison shows that the chemical route for the water formation is the major actor in the game, whilst other differences in the two models, including the physical evolution treatment, play a minor role in the water deuteration. However, we should point out that the O+H$_2$ reaction cannot occur at low temperatures given its endothermicity. Therefore, it is unlikely that water deuteration shows a such strong temperature dependence.

\section{Discussion} \label{sec: discussion}

In the previous paragraphs, we have shown that the deuteration of water and other molecules strongly depends on the H$_2$ opr and the total density, but also on the visual extinction and the temperature where solid molecules are formed. 
The high deuteration of water (HDO/H$_2$O $> 1$ \% and D$_2$O/H$_2$O $> 0.01$ \%) observed towards the low-mass protostar IRAS 16293 \citep{Coutens2012} can only be reproduced with an H$_2$ opr that is lower than $3 \times 10^{-4}$, suggesting that most of deuterated water is formed at low H$_2$ opr. 
The ortho-para ratio reaches its steady-state value on a timescale proportional to the total density, of about $10^7$ yr at $n_H =10^4$ cm$^{-3}$ \citep{Flower2006}. 
If the H$_2$ molecule has an initial opr value of 3 upon its formation on grain surfaces, the high deuteration observed around IRAS 16293 would suggest that water observed in this envelope has been formed in an ''old" molecular gas, i.e. a gas old enough to show a low H$_2$ opr at the moment of the formation of interstellar ices. 
However, it is possible that the ortho-to-para conversion also occurs on grain surfaces \citep{Lebourlot2000} but with an uncertain rate \citep[see][]{Sugimoto2011, Chehrouri2011, Hama2012}. The decrease in the H$_2$ opr could occur, therefore, faster and the molecular cloud age of $10^7$ yr deduced from the estimate by \citet{Flower2006} should only be used as an upper limit. 

The observed deuterium fractionation of water is only reproduced for dark conditions (high visual extinctions and low temperatures if $n_H = 10^4$ cm$^{-3}$ or higher densities). Therefore, although IR observations of ices show that water ice starts to form at low visual extinctions, the deuterated water observed in IRAS 16293 should instead be formed in darker regions. 
Consequently, H$_2$O would be present in the inner part of ice mantles, whilst most of HDO and D$_2$O molecules should be located in the outer layers. 
A physical evolution, modelling the accumulation of matter from diffuse molecular clouds to dense cores, would allow us to directly confirm this result. 
The high observed deuteration of water also confirms that water is mostly formed from reactions involving atomic H and D and not by molecular hydrogen (see Sec. \ref{comp_models}).

\begin{figure}[htp]
\centering
\includegraphics[width=88mm]{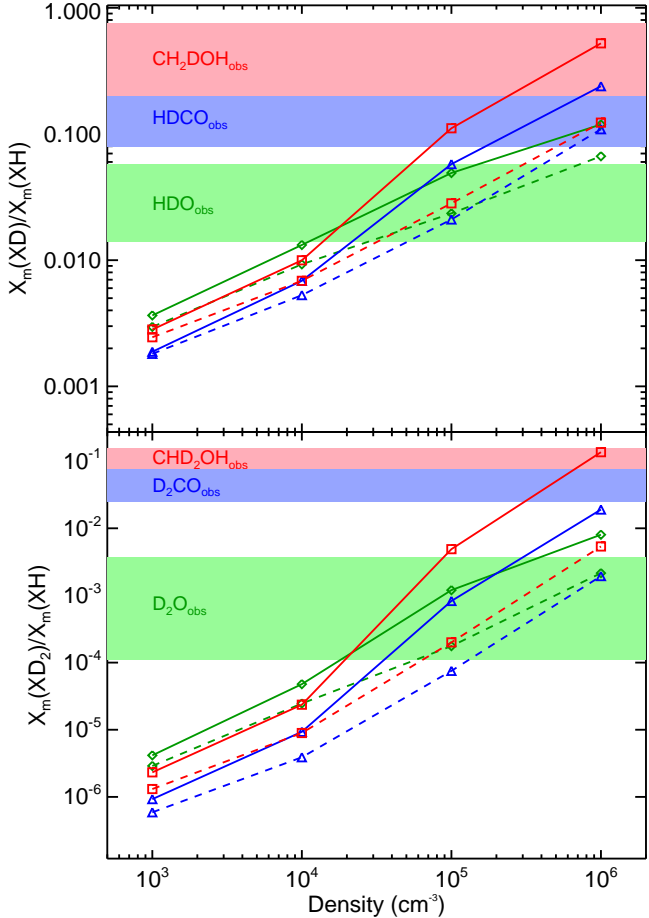}
\caption{Averaged deuterations of water (green diamonds), formaldehyde  (blue triangles), and methanol (red squares) at $3 \times 10^5$ yr (representative of a typical prestellar core age) as function of the total density (abscissa) and temperature (10 K: solid, 20 K: dashed) considering an H$_2$ opr of $3 \times 10^{-6}$ and $A_V = 10$ mag and considering grain surface parameters as free. Solid boxes refer to observations by \citet{Coutens2012} towards IRAS 16293 for water, and by \citet{Parise2006} (and references therein) towards a sample of low-mass protostars for formaldehyde and methanol.}
\label{obs_mod}
\end{figure}

Observations of water, formaldehyde, and methanol vapours towards low-mass protostars show that these species have different deuterium fractionations. Figure \ref{obs_mod} overplots the observed deuteration towards IRAS 16293 with mean theoretical deuteration values reached at $t = 3 \times 10^5$ yr \citep[i.e., the upper limit of the age of prestellar cores, see][]{Bergin2007} as function of the density, for two temperatures, a low H$_2$ opr of $3 \times 10^{-6}$ and only considering dark regions ($A_v = 10$ mag). 
The comparison of our predictions with observations shows that \\
i) our model reproduces the observed HDO/H$_2$O and D$_2$O/H$_2$O ratios for densities between $1 \times 10^4$ and $1 \times 10^5$ cm$^{-3}$ at 10 K and between and $2 \times 10^4$ and $3 \times 10^5$ cm$^{-3}$ at 20 K. Therefore, water deuteration is reproduced within a wide range of physical conditions representative of molecular clouds but not in the too-translucent cloud phase with too low density, too low visual extinctions and too high temperatures. \\
ii) the observed HDCO/H$_2$CO can be reproduced at higher densities ($3 \times 10^5 - 10^6$ cm$^{-3}$) and lower temperatures ($\sim 10$ K) as seen in the central regions of prestellar cores. The D$_2$CO/H$_2$CO ratio is reproduced at densities higher than $10^6$ cm$^{-3}$ (at $\sim 5 \times 10^6$ cm$^{-3}$, TCK12b). \\
iii) methanol deuteration (CH$_2$DOH/CH$_3$OH and CHD$_2$OH/CH$_3$OH ratios) proceeds in the outer parts of grain mantles when the prestellar core condensation reached high densities ($> 5 \times 10^5$ cm$^{-3}$) and low temperatures (10 K). 

In summary, the difference in the deuterium fractionation of molecules seen in IRAS 16293 can be explained by different periods of formation. Water is mainly formed first in regions showing intermediate densities representative of molecular clouds ($n_H = 10^4 - 10^5$ cm$^{-3}$ where the CO depletion is limited. Formaldehyde and methanol are formed subsequently, at higher CO depletions when the gas is denser and colder. These predictions are in good agreement with infrared observations of interstellar ices presented in the introduction which also show that water ice is formed first, at low visual extinctions, whilst solid methanol forms later on at visual extinctions higher than 15 mag \citep{Whittet2011}.

More generally, the deuterium fractionation of any species, believed to be formed mainly in ices, can be used as a tracer to estimate the physical conditions at the moment of its formation. As shown in Fig. \ref{obs_mod}, the D/H ratio increases with increasing density and with decreasing temperature. A low deuteration, similar to the deuteration of water, implies that the solid species is mainly formed in regions with relatively low densities, even if its deuterated isotopologues could be formed mainly in a later phase. In contrast, a high deuteration level suggests that the solid species is mostly formed in the dense and cold phase where the deuteration in the gas-phase (and in particular the atomic D/H ratio) is high. Extremely high deuteration, such as methanol deuteration, would also imply the existence of abstraction reactions or other processes enhancing the formation of deuterated isotopologues with respect to the main isotopologue (TCK12b). 
H/D exchange reactions between solid methanol and water, for example, can also lead to a selective deuteration of functional groups \citep{Ratajczak2009}.


\section{Conclusions}

We have presented a comprehensive study of the formation and the deuteration of interstellar water ice carried out with our astrochemical model GRAINOBLE. In addition to the multilayer formation of grain mantles presented in a first paper (TCK12a) and the use of abstraction reactions for the deuteration of formaldehyde and methanol (TCK12b), we introduced a better treatment for computing the transmission probabilities of surface reactions based on the Eckart model. We also considered wavelength-dependent UV photodesorption of ices following molecular dynamics (MD) simulations and new experimental works.

The main results of this work are the following: 

1) Our model reproduces the abundance of water ice and the visual extinction threshold observed by \citet{Whittet1988}, as well as the different ice components seen in different molecular clouds regions.

2) Water deuteration strongly depends on the ortho-to-para ratio of H$_2$ and the total density, but also, even though more weakly, on the gas and grain temperatures and on the visual extinction. 

3) The deuteration of water observed towards the low-mass protostar IRAS 16293 can only be reproduced by considering a H$_2$ opr lower than $3 \times 10^{-4}$ and a total density between  $8 \times 10^3$ and $5 \times 10^5$ cm$^{-3}$. {If a low density ($10^4$ cm$^{-3}$) is considered, a low temperature (10 K) and a visual extinction higher than 4 mag are necessary. Dark regions (high densities or high visual extinctions) are, therefore, needed to reproduce the observations.}

{4) Comparison between the observed deuteration of water, formaldehyde, and methanol and our theoretical predictions allowed us to propose the following scenario. Water ice is formed first within relatively wide ranges of physical conditions: H$_2$O is allowed to start its formation at low densities and low visual extinctions, but HDO and D$_2$O are instead formed in darker (higher $n_{\textrm{H}}$ and/or $A_V$) regions. Formaldehyde and then methanol are mainly formed subsequently, in the dense and cold prestellar cores.}

5) Deuterium fractionation can be used to estimate the values of the density and temperature at the moment of formation of solid species because of its sensitivity to the physical conditions. Low deuteration, similar to water deuteration, implies that the species is formed in a wide range of physical conditions, whilst a high deuteration suggests an efficient formation in the centre of dense and cold prestellar cores.

With this work, we have explored the influence of a wide range of parameters values on the formation of deuterated ices. 
Now that the influence of the various parameters is clarified, our next step will be a more sophisticated model, based on a realistic physical evolution of a cloud, in a forecoming paper.

\begin{acknowledgements}

The authors would like to thank Naoki Watanabe, Yasuhiro Oba, and Tetsuya Hama for fruitful discussions and for their useful comments about the surface reactions occurring in cold conditions, and Guillaume Pineau des Forets for discussions about H$_2$ photodissociation and gas-phase abundance of deuterium. 
V. T. would like to thank Alexandre Faure for discussions about the H$_2$ ortho/para ratio and a careful reading of the manuscript, and Edith Fayolle for providing useful data on photolysis.
This work has been supported by l\textquoteright Agence Nationale pour la Recherche (ANR), France (project FORCOMS, contracts ANR-08-BLAN-022). A.L.S. and C.C. acknowledge funding from the CNES (Centre National d’E ́tudes Spatiales).
The computations of the model grid presented in this work were performed  at the Service Commun de Calcul Intensif de l'Observatoire de Grenoble (SCCI). 
Quantum chemical calculations were performed thanks to the HPC resources of CINES under the allocation 2012-088620 made by GENCI (Grand Equipement National de Calcul Intensif). 
Some kinetic data we used have been downloaded from the online database KIDA (KInetic Database for Astrochemistry, http://kida.obs.u-bordeaux1.fr).

  \end{acknowledgements}

\bibliographystyle{aa}

\bibliography{bib3}

 \newpage
\appendix

\section{Transmission probability computations}

\subsection{Eckart model}

We compute the transmission probabilities of all the reactions involved in the water-producing and the methanol-producing networks by using the Eckart model \citep{Eckart1930, Johnston1962}. In this approach, an approximate potential energy surface (PES) is fitted as a function of the zero point energies (ZPEs) of the stationary points. The parameters needed for computing the transmission probability are the zero-point-corrected barrier heights of the forward and reverse reactions $V_f$ and $V_r$, the frequency of the imaginary mode of the transition state $\nu_S$, and the reduced mass of the reactants $\mu$. 

The Eckart potential can be parametrised as
\begin{equation}
U_x = \frac{A \exp \left( {\frac{x-x_0} {l}}\right) } {1+\exp \left( {\frac{x-x_0}{l}}\right)} + 
\frac{B \exp \left({\frac{x-x_0}{l}}\right)} {\left(1+\exp \left({\frac{x-x_0}{l}}\right) \right)^2}
\end{equation}
where
\begin{equation} A = V_f - V_r \end{equation}
\begin{equation} B = \left(\sqrt{V_f} +\sqrt{V_r}\right)^2 \end{equation}
\begin{equation} l = \frac{2 \pi}{|\nu_S|} \sqrt{\frac{2}{\mu}} \left(\frac{1}{\sqrt{V_f}} +\frac{1}{\sqrt{V_r}}\right)^{-1}. \end{equation}

Once one has fitted the potential then the transmission probability, $P_{r}$, may be calculated using
\begin{equation}
P_r = \frac{\cosh(\alpha+\beta) - \cosh(\alpha-\beta)}{\cosh(\alpha+\beta) + \cosh(\delta)}
\end{equation}
where
\begin{equation} \alpha = \frac{4 \pi}{| \nu_S |} \left(\frac{1}{\sqrt{V_f}} + \frac{1}{\sqrt{V_r}}\right)^{-1} \sqrt{E} \end{equation}
\begin{equation} \beta = \frac{8 \pi^2}{h | \nu_S |} \left(\frac{1}{\sqrt{V_f}} + \frac{1}{\sqrt{V_r}}\right)^{-1} \sqrt{E - V_f + V_r} \end{equation}
\begin{equation} \delta = 4 \pi \sqrt{\frac{V_f V_r 2 \pi}{(h |\nu_S |)^2}-\frac{1}{16}}. \end{equation}
When the reactants are considered as excited by their formation involving an exothermic reaction, $E$ refers to the excess energy of this reaction of formation. Otherwise, $E$ is the thermal energy of the particles. 

The Eckart model provides a significant improvement over square barriers but it can underpredict (or overpredict) the transmission probabilities of some reactions at low temperatures compared to more exact methods \citep[see][]{Peters2011}. However, given the number of surface reactions considered in this work, exact quantum chemical computations for all reactions are not feasible.

\subsection{CO$_2$ formation}

Based on the experimental work by \citet{Oba2010} who studied the formation of CO$_2$ from CO and OH, CO$_2$ is thought to be formed via the pathway
\begin{equation} \textrm{CO} + \textrm{OH} \rightarrow \textrm{trans-HOCO} \label{CO2_2_1} \end{equation}
\begin{equation} \textrm{trans-HOCO} \rightarrow \textrm{cis-HOCO} \label{CO2_2_2} \end{equation}
\begin{equation} \textrm{cis-HOCO} \rightarrow \textrm{CO}_2 + \textrm{H}. \label{CO2_2_3} \end{equation}
The electronic energy of the intermediate radicals (t-HOCO, c-HOCO) and the products (CO$_2$ + H) are lower than for the reactants (CO + OH). Reactions involving t-HOCO and c-HOCO possess high activation barriers and/or are endothermic \citep{Yu2001}. 
Therefore, t-HOCO radicals continue to react only if their excess energy released by the chemical energy of reaction (\ref{CO2_2_1}) is sufficient to overcome the high activation barriers of reactions (\ref{CO2_2_2}) and (\ref{CO2_2_3}). 
As in \citet{Goumans2008}, HOCO radicals can also react with H atoms via barrierless reactions to form three pairs of products: H$_2$ + CO$_2$, H$_2$O + CO, or HCOOH. The lack of more quantitative data leads us to assume a branching ratio of one to three for every product pair.

In fact, the energy released by reaction (\ref{CO2_2_1}) absorbed by HOCO radicals can be transferred to the surface before reacting. However, the transfer rate of the chemical energy to the surface is very uncertain. The absence of formic acid HCOOH and the low abundance of HOCO radicals observed in the experiments of \citet{Oba2010} suggest that CO$_2$ is readily formed from excited HOCO molecules. Therefore, most HOCO radicals continue to react before relaxing to their stable state. 
To reproduce these experiments, we assume that 99 \% of HOCO radicals are sufficiently excited to form CO$_2$, whilst 1\% of them are stabilized. 

\citet{Goumans2010} performed gas-phase O-CO potential energy surface calculations, showing that O and CO can form a van der Waals complex, allowing O atoms to stay bound to CO for a long time. 
Following these results, we consider that O atoms that meet CO molecules (via direct accretion from gas-phase or surface diffusion) form a loosely bound O...CO complex. The H atoms that meet these O..CO complexes react via a barrierless reaction to form an excited HO...CO* complex. If the time for energy transfer to the surface is long enough, the complex can yield OH + CO, or t-HOCO* radical via barrierless reactions. Otherwise, the complex forms the t-HOCO radical through quantum tunneling. We consider that 99\% of HO...CO* complexes continue to react without activation barriers.

Appendix B lists all the reactions involved in the formation of CO$_2$ with their corresponding activation barriers and transmission probabilities.



\subsection{Quantum chemical calculations}

The OH + H$_2$ reaction system has been theoretically studied by \citet{Nguyen2011}. These authors have computed the forward and reverse reactions, the imaginary frequency of the transition states, and the rate constants of the eight reactions involving H$_2$, HD, D$_2$, OH, and OD using the semiclassical transition-state theory (SCTST). 

The H$_2$O$_2$ + H reaction has been studied by \citet{Koussa2006} and \citet{Ellingson2007}. However, data for the reactions involving deuterated isotopologues was not available. Therefore, to obtain the data required for the model, quantum chemistry calculations have been conducted with the Gaussian 09 \citep{Gaussian2009} program. For these calculations, the PBE0 \citep{Perdew1996,Perdew1997,Adamo1999} functional and the aug-cc-pVTZ \citep{Dunning1989,Kendall1992} basis set were used because this combination produced results that are in good agreement with the experimentally determined values for the process in gas-phase \citep[][who measured an activation barrier of 4.6 kcal/mol = 2300 K]{Klemm1975} .

The CO + H reaction has been theoretically studied by several authors \citep{Woon2002, Andersson2011, Peters2012}. We decided to use the work of \citet{Peters2012} because they used the most accurate methodology and obtained a value for the activation energy for the formation of HCO that best agrees the gas-phase experiment \citep{Wang1973}. The transmission probabilities of all the reactions producing deuterated formaldehyde and methanol are computed from this reaction and relative rates measured by \citet{Nagaoka2007} and \citet{Hidaka2009} or deduced by TCK12b.

The formation of carbon dioxide includes two reactions having an activation barrier. 
Quantum calculations of \citet{Talbi2006, Goumans2008}, and \citet{Goumans2010} showed that reaction (\ref{CO2_3}) has an activation barrier of 2500 - 3000 K, leading to a transmission probability of $5 \times 10^{-23}$ \citep{Goumans2010, Garrod2011}. Assuming an activation energy of 2500 K and considering a square barrier, we reproduce this transmission probability with a barrier width of 0.8 $\AA$. 
The transmission probabilities of the reaction pathways involved in reaction (\ref{CO2_2}), and including HOCO radicals, the van der Waals complex HO...CO, and their deuterated isotopologues, were deduced from the potential energy surface computed by \citet{Yu2001}.  These authors computed the stationary points of the potential energy surface of this reaction using an extrapolated full coupled cluster/complete basis set (FCC/CBS) method. 

\section{List of grain surface reactions}

\addtocounter{table}{1}

\longtab{1}{\begin{center}
\begin{longtable}{c c c c c c c c c}
\caption{List of grain surface chemical reactions considered in this work along with their activation barriers and transmission probabilities.} \\
\hline
\hline
Reaction	&		&		&		&		&		&		&	$E_a$ (K)	&	P$_r$	\\
\hline																	
H	&	+	&	H	&	$\rightarrow$	&	(o)H$_2$	&		&		&	0	&	0.75	\\
H	&	+	&	H	&	$\rightarrow$	&	(p)H$_2$	&		&		&	0	&	0.25	\\
H	&	+	&	D	&	$\rightarrow$	&	  HD	&		&		&	0	&	1	\\
D	&	+	&	D	&	$\rightarrow$	&	  (o)D$_2$	&		&		&	0	&	0.66	\\
D	&	+	&	D	&	$\rightarrow$	&	  (p)D$_2$	&		&		&	0	&	0.33	\\
\hline																						
CO	&	+	&	   H	&	$\rightarrow$	&	  HCO	&		&		&	1763	&	1.92(-07)	\\
CO	&	+	&	   D	&	$\rightarrow$	&	  DCO	&		&		&	/	&	1.92(-08)	\\
\hline																	
HCO	&	+	&	  H	&	$\rightarrow$	&	  H$_2$CO	&		&		&	0	&	1	\\
HCO	&	+	&	  D	&	$\rightarrow$	&	  HDCO	&		&		&	0	&	1	\\
DCO	&	+	&	  H	&	$\rightarrow$	&	  HDCO	&		&		&	0	&	1	\\
DCO	&	+	&	  D	&	$\rightarrow$	&	  D$_2$CO	&		&		&	0	&	1	\\
\hline																	
H$_2$CO	&	+	&	 H	&	$\rightarrow$	&	  CH$_3$O	&		&		&	/	&	9.60(-08)	\\
H$_2$CO	&	+	&	 D	&	$\rightarrow$	&	  CH$_2$DO	&		&		&	/	&	9.60(-09)	\\
H$_2$CO	&	+	&	 D	&	$\rightarrow$	&	  HCO	&	+	&	  HD	&	/	&	9.31(-08)	\\
H$_2$CO	&	+	&	 D	&	$\rightarrow$	&	  HDCO	&	+	&	 H	&	/	&	9.31(-08)	\\
HDCO	&	+	&	 H	&	$\rightarrow$	&	  CH$_2$DO	&		&		&	/	&	1.11(-07)	\\
HDCO	&	+	&	 D	&	$\rightarrow$	&	  CHD$_2$O	&		&		&	/	&	1.11(-07)	\\
HDCO	&	+	&	 H	&	$\rightarrow$	&	  HCO	&	+	&	  HD	&	/	&	1.54(-07)	\\
HDCO	&	+	&	 D	&	$\rightarrow$	&	  DCO	&	+	&	  HD	&	/	&	9.31(-08)	\\
HDCO	&	+	&	 D	&	$\rightarrow$	&	  D$_2$CO	&	+	&	 H	&	/	&	9.31(-08)	\\
D$_2$CO	&	+	&	 H	&	$\rightarrow$	&	  CHD$_2$O	&		&		&	/	&	1.27(-07)	\\
D$_2$CO	&	+	&	 D	&	$\rightarrow$	&	  CD$_3$O	&		&		&	/	&	1.27(-08)	\\
D$_2$CO	&	+	&	 H	&	$\rightarrow$	&	  DCO	&	+	&	  HD	&	/	&	7.30(-08)	\\
\hline																	
CH$_3$O	&	+	&	 H	&	$\rightarrow$	&	  CH$_3$OH	&		&		&	0	&	1	\\
CH$_3$O	&	+	&	 D	&	$\rightarrow$	&	  CH$_3$OD	&		&		&	0	&	1	\\
CH$_2$OH	&	+	&	H	&	$\rightarrow$	&	  CH$_3$OH	&		&		&	0	&	1	\\
CH$_2$OH	&	+	&	D	&	$\rightarrow$	&	  CH$_2$DOH	&		&		&	0	&	1	\\
CH$_2$OD	&	+	&	H	&	$\rightarrow$	&	  CH$_3$OD	&		&		&	0	&	1	\\
CH$_2$OD	&	+	&	D	&	$\rightarrow$	&	  CH$_2$DOD	&		&		&	0	&	1	\\
CH$_2$DO	&	+	&	H	&	$\rightarrow$	&	  CH$_2$DOH	&		&		&	0	&	1	\\
CH$_2$DO	&	+	&	D	&	$\rightarrow$	&	  CH$_2$DOD	&		&		&	0	&	1	\\
CHDOH	&	+	&	H	&	$\rightarrow$	&	  CH$_2$DOH	&		&		&	0	&	1	\\
CHDOH	&	+	&	D	&	$\rightarrow$	&	  CHD$_2$OH	&		&		&	0	&	1	\\
CHDOD	&	+	&	H	&	$\rightarrow$	&	  CH$_2$DOD	&		&		&	0	&	1	\\
CHDOD	&	+	&	D	&	$\rightarrow$	&	  CHD$_2$OD	&		&		&	0	&	1	\\
CHD$_2$O	&	+	&	H	&	$\rightarrow$	&	  CHD$_2$OH	&		&		&	0	&	1	\\
CHD$_2$O	&	+	&	D	&	$\rightarrow$	&	  CHD$_2$OD	&		&		&	0	&	1	\\
CD$_2$OH	&	+	&	H	&	$\rightarrow$	&	  CHD$_2$OH	&		&		&	/	&	1	\\
CD$_2$OH	&	+	&	D	&	$\rightarrow$	&	  CD$_3$OH	&		&		&	0	&	1	\\
CD$_2$OD	&	+	&	H	&	$\rightarrow$	&	  CHD$_2$OD	&		&		&	0	&	1	\\
CD$_2$OD	&	+	&	D	&	$\rightarrow$	&	  CD$_3$OD	&		&		&	0	&	1	\\
CD$_3$O	&	+	&	 H	&	$\rightarrow$	&	  CD$_3$OH	&		&		&	0	&	1	\\
CD$_3$O	&	+	&	 D	&	$\rightarrow$	&	  CD$_3$OD	&		&		&	0	&	1	\\
\hline																	
CH$_3$OH	&	+	&	D	&	$\rightarrow$	&	  CH$_2$OH	&	+	&	HD	&	/	&	2.88(-07)	\\
CH$_2$DOH	&	+	&	   D	&	$\rightarrow$	&	  CHDOH	&	+	&	HD	&	/	&	1.92(-07)	\\
CHD$_2$OH	&	+	&	   D	&	$\rightarrow$	&	  CD$_2$OH	&	+	&	HD	&	/	&	1.50(-07)	\\
CH$_3$OD	&	+	&	D	&	$\rightarrow$	&	  CH$_2$OD	&	+	&	HD	&	/	&	2.88(-07)	\\
CH$_2$DOD 	&	+	&	  D	&	$\rightarrow$	&	  CHDOD	&	+	&	HD	&	/	&	1.92(-07)	\\
CHD$_2$OD  	&	+	&	 D	&	$\rightarrow$	&	  CD$_2$OD	&	+	&	HD	&	/	&	1.50(-07)	\\
\hline																	
O	&	+	&	O	&	$\rightarrow$	&	  O$_2$	&		&		&	0	&	1	\\
O$_2$	&	+	&	   O	&	$\rightarrow$	&	  O$_3$	&		&		&	0	&	1	\\
\hline																	
O	&	+	&	H	&	$\rightarrow$	&	  OH	&		&		&	0	&	1	\\
O	&	+	&	D	&	$\rightarrow$	&	  OD	&		&		&	0	&	1	\\
\hline																	
OH	&	+	&	   H	&	$\rightarrow$	&	  H$_2$O	&		&		&	0	&	1	\\
OH	&	+	&	   D	&	$\rightarrow$	&	  HDO	&		&		&	0	&	1	\\
OD	&	+	&	   H	&	$\rightarrow$	&	  HDO	&		&		&	0	&	1	\\
OD	&	+	&	   D	&	$\rightarrow$	&	  D$_2$O	&		&		&	0	&	1	\\
OH	&	+	&	   OH	&	$\rightarrow$	&	 H$_2$O$_2$	&		&		&	0 (R= 0.8)	&	1	\\
OD	&	+	&	   OH	&	$\rightarrow$	&	 HDO$_2$	&		&		&	0 (R= 0.8)	&	1	\\
OD	&	+	&	   OD	&	$\rightarrow$	&	 D$_2$O$_2$	&		&		&	0 (R= 0.8)	&	1	\\
OH	&	+	&	   OH	&	$\rightarrow$	&	 H$_2$O	&	+	&	  O	&	0 (R=0.2)	&	1	\\
OD	&	+	&	   OH	&	$\rightarrow$	&	 HDO	&	+	&	  O	&	0 (R=0.2)	&	1	\\
OD	&	+	&	   OD	&	$\rightarrow$	&	 D$_2$O	&	+	&	  O	&	0 (R=0.2)	&	1	\\
OH	&	+	&	H$_2$	&	$\rightarrow$	&	H$_2$O	&	+	&	  H	&	2935	&	4.07(-07)	\\
OD	&	+	&	H$_2$	&	$\rightarrow$	&	HDO	&	+	&	  H	&	2855	&	3.62(-07)	\\
OD	&	+	&	   HD	&	$\rightarrow$	&	 D$_2$O	&	+	&	  H	&	3051	&	1.00(-09)	\\
OD	&	+	&	   HD	&	$\rightarrow$	&	 HDO	&	+	&	  D	&	3026	&	8.07(-10)	\\
OH	&	+	&	   HD	&	$\rightarrow$	&	 H$_2$O	&	+	&	  D	&	2789	&	8.74(-07)	\\
OH	&	+	&	   HD	&	$\rightarrow$	&	 HDO	&	+	&	  H	&	2900	&	2.81(-09)	\\
OH	&	+	&	D$_2$	&	$\rightarrow$	&	HDO	&	+	&	  D	&	2703	&	7.99(-07)	\\
OD	&	+	&	D$_2$	&	$\rightarrow$	&	D$_2$O	&	+	&	  D	&	2870	&	2.26(-09)	\\
\hline																	
O$_2$	&	+	&	   H	&	$\rightarrow$	&	  HO$_2$	&		&		&	0	&	1	\\
O$_2$	&	+	&	   D	&	$\rightarrow$	&	  DO$_2$	&		&		&	0	&	1	\\
\hline																	
HO$_2$	&	+	&	  H	&	$\rightarrow$	&	  H$_2$O$_2$	&		&		&	0	&	1	\\
HO$_2$	&	+	&	  D	&	$\rightarrow$	&	  HDO$_2$	&		&		&	0	&	1	\\
DO$_2$	&	+	&	  H	&	$\rightarrow$	&	  HDO$_2$	&		&		&	0	&	1	\\
DO$_2$	&	+	&	  D	&	$\rightarrow$	&	  D$_2$O$_2$	&		&		&	0	&	1	\\
\hline																	
H$_2$O$_2$	&	+	&	 H	&	$\rightarrow$	&	  H$_2$O	&	+	&	  OH	&	2508	&	1.37(-07)	\\
H$_2$O$_2$	&	+	&	 D	&	$\rightarrow$	&	  HDO	&	+	&	  OH	&	2355	&	5.54(-09)	\\
HDO$_2$	&	+	&	 H	&	$\rightarrow$	&	  HDO	&	+	&	  OH	&	2523	&	1.23(-07)	\\
HDO$_2$	&	+	&	 H	&	$\rightarrow$	&	  H$_2$O	&	+	&	  OD	&	2524	&	1.22(-07)	\\
HDO$_2$	&	+	&	 D	&	$\rightarrow$	&	  D$_2$O	&	+	&	  OH	&	2369	&	5.28(-09)	\\
HDO$_2$	&	+	&	 D	&	$\rightarrow$	&	  HDO	&	+	&	  OD	&	2367	&	5.29(-09)	\\
D$_2$O$_2$	&	+	&	 H	&	$\rightarrow$	&	  D$_2$O	&	+	&	  OH	&	2540	&	1.08(-07)	\\
D$_2$O$_2$	&	+	&	 D	&	$\rightarrow$	&	  D$_2$O	&	+	&	  OD	&	2384	&	4.28(-09)	\\
\hline																	
O$_3$	&	+	&	   H	&	$\rightarrow$	&	  O$_2$	&	+	&	   OH	&	0	&	1	\\
O$_3$	&	+	&	   D	&	$\rightarrow$	&	  O$_2$	&	+	&	   OD	&	0	&	1	\\
\hline																	
N	&	+	&	H	&	$\rightarrow$	&	  NH	&		&		&	0	&	1	\\
N	&	+	&	D	&	$\rightarrow$	&	  ND	&		&		&	0	&	1	\\
\hline																	
NH	&	+	&	   H	&	$\rightarrow$	&	  NH$_2$	&		&		&	0	&	1	\\
NH	&	+	&	   D	&	$\rightarrow$	&	  NHD	&		&		&	0	&	1	\\
ND	&	+	&	   H	&	$\rightarrow$	&	  NHD	&		&		&	0	&	1	\\
ND	&	+	&	   D	&	$\rightarrow$	&	  ND$_2$	&		&		&	0	&	1	\\
\hline				
NH$_2$	&	+	&	  H	&	$\rightarrow$	&	  NH$_3$	&		&		&	0	&	1	\\
NH$_2$	&	+	&	  D	&	$\rightarrow$	&	  NH$_2$D	&		&		&	0	&	1	\\
NHD	&	+	&	  H	&	$\rightarrow$	&	  NH$_2$D	&		&		&	0	&	1	\\
NHD	&	+	&	  D	&	$\rightarrow$	&	  NHD$_2$	&		&		&	0	&	1	\\
ND$_2$	&	+	&	  H	&	$\rightarrow$	&	  NHD$_2$	&		&		&	0	&	1	\\
ND$_2$	&	+	&	  D	&	$\rightarrow$	&	  ND$_3$	&		&		&	0	&	1	\\
\hline																	
C	&	+	&	H	&	$\rightarrow$	&	  CH	&		&		&	0	&	1	\\
C	&	+	&	D	&	$\rightarrow$	&	  CD	&		&		&	0	&	1	\\
\hline																	
CH	&	+	&	   H	&	$\rightarrow$	&	  CH$_2$	&		&		&	0	&	1	\\
CH	&	+	&	   D	&	$\rightarrow$	&	  CHD	&		&		&	0	&	1	\\
CD	&	+	&	   H	&	$\rightarrow$	&	  CHD	&		&		&	0	&	1	\\
CD	&	+	&	   D	&	$\rightarrow$	&	  CD$_2$	&		&		&	0	&	1	\\
\hline																	
CH$_2$	&	+	&	  H	&	$\rightarrow$	&	  CH$_3$	&		&		&	0	&	1	\\
CH$_2$	&	+	&	  D	&	$\rightarrow$	&	  CH$_2$D	&		&		&	0	&	1	\\
CHD	&	+	&	  H	&	$\rightarrow$	&	  CH$_2$D	&		&		&	0	&	1	\\
CHD	&	+	&	  D	&	$\rightarrow$	&	  CHD$_2$	&		&		&	0	&	1	\\
CD$_2$	&	+	&	  H	&	$\rightarrow$	&	  CHD$_2$	&		&		&	0	&	1	\\
CD$_2$	&	+	&	  D	&	$\rightarrow$	&	  CD$_3$	&		&		&	0	&	1	\\
\hline																	
CH$_3$	&	+	&	  D	&	$\rightarrow$	&	  CH$_3$D	&		&		&	0	&	1	\\
CH$_2$D	&	+	&	 H	&	$\rightarrow$	&	  CH$_3$D	&		&		&	0	&	1	\\
CH$_2$D	&	+	&	 D	&	$\rightarrow$	&	  CH$_2$D$_2$	&		&		&	0	&	1	\\
CHD$_2$	&	+	&	 H	&	$\rightarrow$	&	  CH$_2$D$_2$	&		&		&	0	&	1	\\
CHD$_2$	&	+	&	 D	&	$\rightarrow$	&	  CHD$_3$	&		&		&	0	&	1	\\
CD$_3$	&	+	&	  H	&	$\rightarrow$	&	  CHD$_3$	&		&		&	0	&	1	\\
CD$_3$	&	+	&	  D	&	$\rightarrow$	&	  CD$_4$	&		&		&	0	&	1	\\
\hline																	
HCO	&	+	&	  O	&	$\rightarrow$	&	  CO$_2$	&	+	&	  H	&	0	&	1	\\
DCO	&	+	&	  O	&	$\rightarrow$	&	  CO$_2$	&	+	&	  D	&	0	&	1	\\
\hline																	
CO	&	+	&	   O	&	$\rightarrow$	&	  CO$_2$	&		&		&	2500	&	4.80(-23)	\\
\hline																	
CO	&	+	&	   OH	&	$\rightarrow$	&	 t-HOCO	&		&		&	285	&	3.50(-03)	\\
CO	&	+	&	   OD	&	$\rightarrow$	&	 t-DOCO	&		&		&	121	&	1.63(-01)	\\
CO	&	+	&	   OH	&	$\rightarrow$	&	 c-HOCO	&		&		&	2128	&	5.81(-17)	\\
CO	&	+	&	   OD	&	$\rightarrow$	&	 c-DOCO	&		&		&	1964	&	9.57(-16)	\\
\hline																	
t-HOCO*	&	+	&		&	$\rightarrow$	&	CO	&	+	&	   OH	&	13050	&	1.63(-10)	\\
t-HOCO*	&	+	&		&	$\rightarrow$	&	c-HOCO	&		&		&	4114	&	1	\\
t-HOCO	&	+	&	H	&	$\rightarrow$	&	  CO$_2$	&	+	&	H$_2$	&	0	&	1	\\
t-HOCO	&	+	&	H	&	$\rightarrow$	&	  H$_2$O	&	+	&	  CO	&	0	&	1	\\
t-HOCO	&	+	&	H	&	$\rightarrow$	&	  HCOOH	&		&		&	0	&	1	\\
t-HOCO	&	+	&	D	&	$\rightarrow$	&	  CO$_2$	&	+	&	HD	&	0	&	1	\\
t-HOCO	&	+	&	D	&	$\rightarrow$	&	  HDO	&	+	&	  CO	&	0	&	1	\\
t-HOCO	&	+	&	D	&	$\rightarrow$	&	  HCOOD	&		&		&	0	&	1	\\
\hline																	
c-HOCO*	&	+	&		&	$\rightarrow$	&	t-HOCO	&		&		&	3272	&	1	\\
c-HOCO*	&	+	&		&	$\rightarrow$	&	CO$_2$	&	+	&	  H	&	12440	&	2.63(-01)	\\
c-HOCO	&	+	&		&	$\rightarrow$	&	t-HOCO	&		&		&	3272	&	1.19(-17)	\\
c-HOCO	&	+	&		&	$\rightarrow$	&	CO$_2$	&	+	&	  H	&	12440	&	2.94(-21)	\\
c-HOCO	&	+	&	H	&	$\rightarrow$	&	  CO$_2$	&	+	&	H$_2$	&	0	&	1	\\
c-HOCO	&	+	&	H	&	$\rightarrow$	&	  H$_2$O	&	+	&	  CO	&	0	&	1	\\
c-HOCO	&	+	&	H	&	$\rightarrow$	&	  HCOOH	&		&		&	0	&	1	\\
c-HOCO	&	+	&	D	&	$\rightarrow$	&	  CO$_2$	&	+	&	  HD	&	0	&	1	\\
c-HOCO	&	+	&	D	&	$\rightarrow$	&	  HDO	&	+	&	  CO	&	0	&	1	\\
c-HOCO	&	+	&	D	&	$\rightarrow$	&	  HCOOD	&		&		&	0	&	1	\\
\hline																	
t-DOCO*	&	+	&		&	$\rightarrow$	&	CO	&	+	&	   OD	&	13220	&	9.27(-02)	\\
t-DOCO*	&	+	&		&	$\rightarrow$	&	c-DOCO	&		&		&	4238	&	1	\\
t-DOCO	&	+	&	H	&	$\rightarrow$	&	  CO$_2$	&	+	&	HD	&	0 (R=0.33)	&	1	\\
t-DOCO	&	+	&	H	&	$\rightarrow$	&	HDO	&	+	&	  CO	&	0 (R=0.33)	&	1	\\
t-DOCO	&	+	&	H	&	$\rightarrow$	&	DCOOH	&		&		&	0 (R=0.33)	&	1	\\
t-DOCO	&	+	&	D	&	$\rightarrow$	&	  CO$_2$	&	+	&	 D$_2$	&	0 (R=0.33)	&	1	\\
t-DOCO	&	+	&	D	&	$\rightarrow$	&	  D$_2$O	&	+	&	  CO	&	0 (R=0.33)	&	1	\\
t-DOCO	&	+	&	D	&	$\rightarrow$	&	  DCOOD	&		&		&	0 (R=0.33)	&	1	\\
\hline																	
c-DOCO*	&	+	&		&	$\rightarrow$	&	t-DOCO	&		&		&	3343	&	1	\\
c-DOCO*	&	+	&		&	$\rightarrow$	&	CO$_2$	&	+	&	  D	&	13230	&	6.18(-02)	\\
c-DOCO	&	+	&		&	$\rightarrow$	&	t-DOCO	&		&		&	3343	&	1.51(-22)	\\
c-DOCO	&	+	&	D	&	$\rightarrow$	&	  CO$_2$	&	+	&	D$_2$	&	0 (R=0.33)	&	1	\\
c-DOCO	&	+	&	D	&	$\rightarrow$	&	  D$_2$O	&	+	&	  CO	&	0 (R=0.33)	&	1	\\
c-DOCO	&	+	&	D	&	$\rightarrow$	&	  DCOOD	&		&		&	0 (R=0.33)	&	1	\\
c-DOCO	&	+	&	H	&	$\rightarrow$	&	  CO$_2$	&	+	&	  HD	&	0 (R=0.33)	&	1	\\
c-DOCO	&	+	&	H	&	$\rightarrow$	&	  HDO	&	+	&	  CO	&	0 (R=0.33)	&	1	\\
c-DOCO	&	+	&	H	&	$\rightarrow$	&	  DCOOH	&		&		&	0 (R=0.33)	&	1	\\
\hline																	
O	&	+	&	CO	&	$\rightarrow$	&	 O...CO	&		&		&	0	&	1	\\
 O...CO	&	+	&	  H	&	$\rightarrow$	&	HO...CO	&		&		&	0	&	1	\\
HO...CO	&		&		&	$\rightarrow$	&	t-HOCO	&		&		&	775	&	9.46(-09)	\\
HO...CO	&		&		&	$\rightarrow$	&	c-HOCO	&		&		&	2618	&	8.17(-21)	\\
HO...CO*	&		&		&	$\rightarrow$	&	  t-HOCO	&		&		&	775	&	1	\\
HO...CO*	&		&		&	$\rightarrow$	&	c-HOCO	&		&		&	2618	&	1	\\
DO...CO	&		&		&	$\rightarrow$	&	t-DOCO	&		&		&	727	&	2.97(-18)	\\
DO...CO	&		&		&	$\rightarrow$	&	c-DOCO	&		&		&	2769	&	1.60(-20)	\\
DO...CO*	&		&		&	$\rightarrow$	&	  t-DOCO	&		&		&	727	&	1	\\
DO...CO*	&		&		&	$\rightarrow$	&	c-DOCO	&		&		&	2769	&	1	\\
\hline												
\end{longtable} \end{center}}

\end{document}